# How understanding large language models can inform the use of ChatGPT in physics education

Giulia Polverini and Bor Gregorcic

*Department of Physics and Astronomy, Uppsala University, Box 516, 75120 Uppsala, Sweden*

**Abstract:** The paper aims to fulfil three main functions: (1) to serve as an introduction for the physics education community to the functioning of Large Language Models (LLMs), (2) to present a series of illustrative examples demonstrating how prompt-engineering techniques can impact LLMs performance on conceptual physics tasks and (3) to discuss potential implications of the understanding of LLMs and prompt engineering for physics teaching and learning. We first summarise existing research on the performance of a popular LLM-based chatbot (ChatGPT) on physics tasks. We then give a basic account of how LLMs work, illustrate essential features of their functioning, and discuss their strengths and limitations. Equipped with this knowledge, we discuss some challenges with generating useful output with *ChatGPT-4* in the context of introductory physics, paying special attention to conceptual questions and problems. We then provide a condensed overview of relevant literature on prompt engineering and demonstrate through illustrative examples how selected prompt-engineering techniques can be employed to improve *ChatGPT-4*'s output on conceptual introductory physics problems. Qualitatively studying these examples provides additional insights into ChatGPT's functioning and its utility in physics problem solving. Finally, we consider how insights from the paper can inform the use of LLMs in the teaching and learning of physics.

## I. INTRODUCTION

In late 2022, the public release of a large language model-based chatbot, OpenAI's ChatGPT, introduced a new type of entity into the educational landscape, showing the potential for significant disruptions of educational processes [1]. Recent reports from Sweden, for example, show that around half of high-school students have already used ChatGPT for "cheating" on school assignments [2] and similar findings are emerging from elsewhere [3]. As we come to terms with large language models (LLMs) already being part of the teaching and learning ecosystem in different subjects, including physics, it is important for the community of physics educators to build some understanding of their ways of functioning and what they are capable of. In response to the emerging need for a better understanding of LLMs [1], the aim of this paper is to present some rudimentary concepts concerning LLMs' functioning on a level that can be understood by physics educators and physics education researchers without in-depth knowledge of machine learning. By acquiring some rudimentary understanding of LLMs' functioning, physics educators can better understand these models' inherent strengths and limitations and leverage this understanding to inform their practice.

In section II, we summarise the still relatively sparse but quickly growing body of research on the performance of LLMs on physics tasks, as well as provide a preview of our own findings on the topic. In section III, we first give a simplified technical overview of how LLMs function, leading up to an unpacking of their limitations and the discussion of key differences between LLMs and humans. These insights can help us build a better understanding of the findings reported in the literature on LLMs' performance in physics and more broadly. Insights from section III also lay the ground for section IV, where we demonstrate how our interaction with ChatGPT-4 can be optimised for its performance in physics problem-solving. Through a series of illustrative examples, we showcase two promising approaches to LLM prompting and discuss how insights from the process of designing prompts (also referred to as *prompt engineering*) can contribute to scholarly discussion in PER and adjacent fields of study. Section V offers a broader discussion on the educational implications of understanding LLMs' functionality.

## II. EXISTING RESEARCH ON LLM PERFORMANCE ON PHYSICS TASKS

Even before the public release of the now well-known LLM-powered chatbot ChatGPT, LLMs were studied in different contexts and were found to be well-performing in natural language tasks, including common-sense reasoning, question-answering, and summarisation [4,5]. They were also found to be struggling with tasks that require quantitative reasoning, such as solving mathematics, science, and engineering problems [6,7], and were likely to make up false facts [8]. Since the release of ChatGPT, this field of research has significantly expanded, with scholars in different fields studying ChatGPT's performance in their subject areas [9–12]. Scholars generally find the progress in the development of the technology impressive, but also point out its limited abilities in different areas.

Today, GPT-4 is considered the state-of-the-art LLM, outperforming its predecessor, GPT-3.5, on most academic and professional exams [13–15]. Bubeck et al. [16] provide an early but accurate demonstration of how GPT-4 can solve novel complex tasks on an almost human-like level of performance. However, its level of performance in mathematics has been shown to be quite far from that of experts and it can still fail in some elementary mathematical and common-sense reasoning tasks [13,16], when not augmented by external plugins.

The PER community has also been quick to start exploring how ChatGPT performs on different physics-related tasks. Gregorcic and Pendrill [17] presented a case study in which ChatGPT-3.5's answer to a simple problem about mechanics contains wrong and often contradictory physics reasoning. Drawing on this work, dos Santos [18] proposed the same question to several LLM-powered chatbots (ChatGPT-4, Bing, and Bard), carefully describing the quality of all the outputs. He found that ChatGPT-4 outperforms the other chatbots (including its previous version) by providing a both correct and detailed answer. More importantly, it does not present any inconsistency in its arguments. This work is interesting because it highlights some important differences between typical human reasoning and the chatbot's output. In order to observe these differences, however, one needs to ask the chatbot to answer an open question or to explain how it came to its conclusion.

On the other hand, looking at the overall performance of ChatGPT on a multiple-choice questionnaire, it becomes almost impossible to distinguish between a human and an LLM. Kortemeyer [19] tested ChatGPT-3.5 performance on the Force Concept Inventory (FCI) [20], where it scored comparably to beginner students (a score of 50–60%). These findings are corroborated by West's [21] study. In a successive paper, West [22] highlighted the improvements in performance on FCI that characterise the upgrade from ChatGPT-3.5 to ChatGPT-4. He found that ChatGPT-4 typically scores between 90 and 100% correct on the FCI, which is comparable to the performance of physics experts. West [22] also found that the chatbot's responses on the FCI became more stable. While ChatGPT-3.5 typically provided significantly different responses on consecutive attempts, ChatGPT-4 has displayed a zero median amount of variation in its responses.

Going beyond multiple-choice questionnaires, Yeadon et al. [23] demonstrated that ChatGPT-3.5 would achieve first-class grades on short-form physics essays in a university physics course about the history and philosophy of physics. Kortemeyer [19] challenged ChatGPT-3.5 on different kinds of assessment problems taken from an introductory university physics course. The chatbot performance varied from 46 – 55% of correctness on exams and homework, to 90–93% in programming exercises and clicker questions. A recent study by Yeadon and Hardy [24] assessed ChatGPT-3.5's ability to answer physics exam questions across three different academic levels, from secondary school to introductory university courses. They found that at the introductory university level, ChatGPT-3.5 correctly solved between 30 and 40% of problems.

While the performance of ChatGPT-3.5 and -4 on some of the above-mentioned tasks has been reported to be very good (especially on FCI, essay writing, and programming tasks), even its most recent version (ChatGPT-4) was found to perform quite unreliably on a diverse set of physics examinations (i.e., Foundations of Physics 1 and 2, Theoretical Astrophysics, etc.) [25]. These findings suggest that the relatively unimpressive performance that Kortemeyer [19] and Yeadon and colleagues [24,25] found with ChatGPT-3.5 remains an issue with ChatGPT-4. Notable exceptions here are programming tasks, on which ChatGPT-3.5 already performed very well, and FCI tasks, on which ChatGPT-4 showed significant improvement [22]. Based on these findings, Yeadon and Halliday [25] argue that in most cases, only the most poorly performing students would benefit from their exams being solved by ChatGPT-3.5 or -4. They conclude that the chatbot does not yet pose a significant threat to the fidelity of physics assessment exams similar to those used in their study.

In our own investigation, we did not opt for FCI tasks, since they have already been studied, and there is a significant risk that they were also present extensively in the latest model's training data. Most of the tasks that we used in our investigations presented in this paper were original, inspired by end-of-chapter problems from Etkina, Planinsic and Van Heuvelen's *College Physics: Explore and apply* textbook [26]. The tasks we used were conceptually oriented, with little to no computation required, but often requiring some simple symbolic mathematical reasoning to solve.

Throughout sections III and IV, we present excerpts from our dialogue with ChatGPT-4 that add to the body of research on ChatGPT's performance on physics tasks, further supporting the finding that it is still quite unreliable in solving them. Based on these findings, we could indeed be led to conclude that even the most state-of-the-art LLM-based chatbot (ChatGPT-4) is not up to the task of reliably solving conceptual physics problems. However, as we illustrate in section IV, an LLM's performance (reliability and correctness) on conceptual physics tasks can be significantly improved by applying a relatively basic understanding of how LLMs function and using simple but effective prompt-engineering techniques. We would like to note that the examples presented in this paper were generated in May 2023. Because of the rapid pace of development in the field and constant improvement made to ChatGPT-4, it is possible that the same prompts will not result in similar outputs now. Therefore, the primary role of the dialogue excerpts in the paper is not to serve as data for a comprehensive assessment of ChatGPT-4's performance or to predict its future performance on conceptual physics tasks. Instead, we use the examples to support our discussion of (1) how LLMs function and (2) how prompt engineering can improve LLMs' performance on conceptual physics tasks.

## III. HOW LARGE LANGUAGE MODELS WORK

### A. A simplified overview of the fundamental principles of LLM functioning

In 2022, we have witnessed the public release and mass uptake of several modern artificial intelligence programs utilising machine learning to process large amounts of data, including images, audio recordings and text. Furthermore, in addition to the ability to process data, these programs can generate outputs that resemble the data that they were fed. For this reason, they are often referred to as *generative AI*. One subset of these so-called generative AI programs is *large language models* (LLMs). LLMs were designed to process and generate natural language but have also proven helpful for processing and generating programming languages and even symbolic languages, such as mathematics. OpenAI's GPT series and its implementation in the chatbot application ChatGPT is the most prominent example of a broadly accessible LLM application. The GPT acronym stands for *generative pre-trained transformer* [27] and is nowadays often used as a synonym for LLMs. To set the stage for a rudimentary explanation of how LLMs work, we briefly unpack the GPT acronym. For an explanation of the difference between GPT and ChatGPT, see the following subsection (III.B).

*Generative*. Early language models were mostly designed to encode input. The LLMs that have become known to the public in the last year also have the ability to generate text as output. This ability is ostensibly what made them interesting to the broader public. OpenAI's GPT-3 and GPT-4, for example, are both generative models that output text based on their training and a user-provided textual prompt.

*Pre-trained*. An LLM such as GPT is essentially a neural network with a large number (counted in billions) of learning parameters. Training an LLM involves using machine learning algorithms to process and encode a large corpus of text. The training process typically requires a lot of computing resources and results in a *pre-trained* neural network, which is not changed further as the model is used by the end user. Two crucial factors for the success and broad applicability of LLMs, such as the GPT series, are their efficient architecture and a sufficient quantity and quality[1] of the training data [28]. While OpenAI did not openly declare what textual datasets they used to train GPT-4, their size is speculated to be roughly comparable to that of the entire internet.

*Transformer*. Before 2018, LLM training was difficult to do at large scales because of the inefficiency of the machine-learning approaches and the need for manually processing data before it could be fed into the pre-training process. In 2018, the *transformer* architecture significantly simplified the pre-training process by making it possible for LLMs to be trained on essentially unprocessed text, dramatically decreasing the difficulty of model training on truly large datasets. The transformer architecture was a result of developmental work at Google for the purposes of machine translation. It makes use of special mechanisms ("attention" and "self-attention") which allow the model to successfully capture complex relations between words in a given input. For a more technical discussion of the transformer architecture, see [29].

The LLM training process is done by mechanisms of deep learning, where a computer program automatically processes a large amount of language data and encodes its regularities into a large neural network. Using some additional steps, this neutral network can also be used to generate natural language output. A familiar tool that illustrates the way LLMs generate text is that of an auto-complete function used in popular online e-mail services. Given some initial input (such as the beginning of a sentence), the model produces the output based on what is essentially an advanced form of statistical inference. The inference is done by running a predictive algorithm based on the pre-trained neural network and the specific textual input given to it on the spot, also referred to as the *prompt*. If we were to provide a somewhat simplified account of the way LLMs generate output, we could say that the LLM always answers the question: "According to your model of the statistics of human language, what words are likely to come next?" [30 p.2].

As a consequence of the functioning principles of LLMs, the output is always produced on the spot as a statistical inference of the most likely words to follow the prompt. Thus, it does not directly reproduce parts of the data it was trained on, even if some sequences of words can appear in the training data with such regularity that the model is likely to reproduce them *verbatim* (e.g., standard formulations of Newton's laws). Still, it is essential to note that this is not because the model has directly copied its output from the learning dataset but because it is statistically a very likely sequence of words. This has several interesting consequences. For example, the LLM's output tends to be of high quality in terms of syntax and richness of vocabulary. On the other hand, the same process of statistical inference can also produce sequences of words which are in their content untrue when checked against real-world information. Such statements are often referred to as "hallucinations in natural language generation" [31] and represent an important challenge to LLMs' usage in different contexts. We further address this issue in section II.C.

To illustrate the working principle of an LLM, we present two example texts generated on OpenAI's *GPT playground* (https://platform.openai.com/playground) – a developer application that gives end users access to their GPT-3 "auto-complete" model, also referred to as "the bare-bones" model, and allows them to modify some of its settings. GPT playground also allows end users to display a ranked list of words that the model deemed to be the most likely choice for continuing the output sentence (see example below). The words that we gave to GPT-3 as input from which to continue the text were: *"In physics, force is"* and *"In law, force is"*, respectively. Fig. 1 shows both outputs and the calculated probabilities for the first noun in the continuation of the sentence generated by GPT. If the LLM is set to always choose the most likely word from the list shown in Fig.1, the output becomes deterministic, always resulting in the same string of words. In LLM applications this is not always desired, so some randomness is often intentionally introduced. The degree of randomness is controlled by the "temperature" setting. A high temperature setting will result in more diverse outputs, with the probability of a word being selected depending both on the "temperature" and a word's statistical weight (percentage value displayed next to it in the list shown in Fig. 1). Note that "temperature" can be controlled by the user in OpenAI playground or in the ChatGPT's application programming interface intended for app developers, but not in ChatGPT, where it is set to a non-zero value.

---

[1] With extremely large models such as GPT-4 it is less clear if data quality plays as important of a role as it did in smaller models.

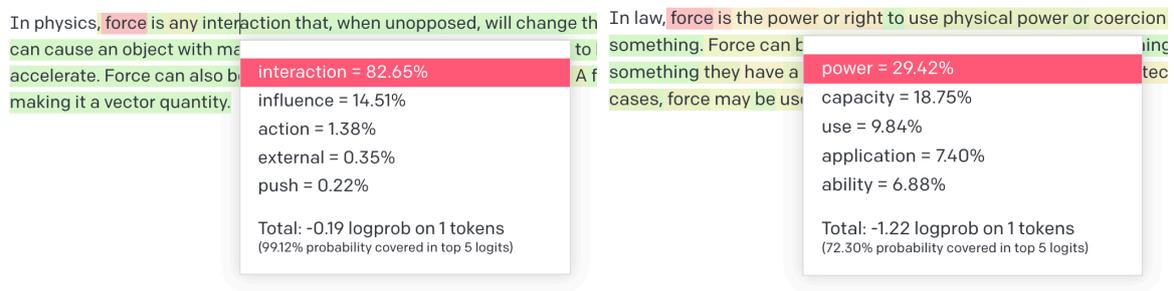

In physics, force is any interaction that, when unopposed, will change the motion of an object. In other words, a force can cause an object with mass to change its velocity (which includes to begin moving from a state of rest), i.e., to accelerate. Force can also be described intuitively as a push or a pull. A force has both magnitude and direction, making it a vector quantity.

In law, force is the power or right to use physical power or coercion against somebody in order to make them do something. Force can be used to compel someone to do something against their will, or to prevent them from doing something they have a right to do. Force can also be used to protect oneself or another person from harm. In some cases, force may be used to protect property or to enforce laws.

**Figure 1**. GPT playground gives end users access to OpenAI's GPT bare-bones model (not ChatGPT), allowing for experimentation and displaying probabilities of words that follow the preceding words. Note that in the physics context the word "interaction" has a much larger statistical weight than the others. This means that even with non-zero "temperature" it will be the most likely choice. On the other hand, in the context of law, the word "power" is not equally dominating. This will result in a more diverse output at the same "temperature".

We can see from the two examples provided in Fig. 1 that GPT's output is context-sensitive [32]. The given "definitions" of force do not bleed into each other but are provided independently in their respective domains (physics and law)[2]. The model responds to a discipline-specific prompt by focusing on a specific disciplinary domain. This will be important for considerations that we bring up later in the paper.

---

[2] This context sensitivity, ensured through the mechanism of self-attention [29], is an important feature of the transformer architecture and significantly contributed to its broad uptake.

## B. From LLMs to chatbots

Most readers will likely be familiar with ChatGPT, which, as we briefly allude to in the previous section, is not quite the same as the GPT models. If ChatGPT were a vehicle, then GPT would be its engine. When an LLM (GPT) is used for a chatbot application (ChatGPT), additional steps need to be taken to provide the users with a good interaction experience. A bare-bones "advanced auto-complete algorithm" model, like GPT, does not yet automatically allow for a dialogue-like interaction[3] or following the user's instructions. As an advanced auto-complete algorithm, an LLM is more likely to continue the instruction than execute what the instructions say. There is a similar problem with answering user questions. Shanahan [30] explains that because answers to questions rarely directly follow the question itself in most data sources, it is unlikely that "auto-completing" the text following the question mark would provide the answer to the question asked. One technique to make it behave like a conversation partner and actually answer the question is embedding a user-posed question into the format of a script of a dialogue between two actors and giving the LLM the task of continuing the dialogue by writing the line of one of the actors[4]. For an illustration of the use of this trick, see section IV.C.3.

In addition to such prompting tricks, going from GPT to ChatGPT involved additional pre-training steps. Following the initial phase of automatised machine learning on large datasets, additional training was done in the form of what is referred to as *reinforcement learning from human and AI feedback.* Reinforcement learning is a process of teaching an AI program the desired behaviour by establishing a reward-seeking system that aims to achieve a given objective. In the case of ChatGPT, this involved humans manually ranking a large number of the model's responses by their desirability, feeding that information into a machine learning algorithm, and using it to fine-tune the model for use as a chatbot. ChatGPT is therefore an application built upon GPT. Creating it involved additional training steps, ensuring that it behaves like a human conversational partner.

In this paper, we use the shorthands ChatGPT-3.5 and ChatGPT-4 to refer to ChatGPT using GPT-3.5 or GPT-4 as its underlying model, respectively. At the time of writing, ChaGPT-3.5 is available as a cost-free service in most western countries, while ChatGPT-4 requires a paid subscription of around 20 USD per month.

## C. Limitations of LLMs

From the previous sections, we can already see that the process through which LLMs predict and generate word sequences does not involve any form of human-like understanding, knowing, thinking or reasoning. A bare-bones LLM such as GPT has no perception of the difficulty of the task we are asking it to solve, nor if the task is meaningful or reasonable. It merely "produces the text that is a statistically good fit, given the starting text" [33 p.684]. There are real downsides to such models, which are in-turn passed down to chatbots. Borji [34] lists eleven categories of possible failures of ChatGPT, including errors in reasoning, math, coding, and factual errors, as well as the display of bias, all carrying potential risks and societal implications. Increasing the amount of learning data used to train the LLM can positively influence the correctness of the outputs, but it seems to do this better in some domains compared to others. For example, Lo [35] shows how ChatGPT achieves outstanding performance in economics and programming tasks while it performs less well in law and medical education. Below, we take a closer look of (1) LLMs' bias, (2) their tendency to generate false information, (3) the prospect of the use of external plugins to remedy some of LLMs' limitations, and (4) LLMs' challenges with coherent argumentation, particularly in physics contexts.

---

[3] The LLM itself includes just the model architecture and pre-trained learning parameters.
[4] This "trick" can provide a first glimpse into the craft of prompt engineering, discussed in section IV.

*1. Bias*

The training dataset on which LLMs are trained often contains different forms of bias. This is not surprising, considering that the data consists of different sources, including the internet, and it is often too large to be actively supervised to eliminate even overtly biased or hateful content. In the training process, the LLM acquires the bias similarly to any other statistical correlation in the training data. Bias is then reflected, for example, by LLMs' tendencies to relate certain traits to specific groups of people, often in a stereotypical way. Researchers have reported that LLMs exhibit bias regarding gender [36,37], race [38,39], religion and caste [40], as well as political views [41]. It is noteworthy that these biases present an important challenge for LLMs' use in education. Learners using LLMs without being aware and critically reflecting on their biases may result in their further propagation. Understanding, measuring and mitigating bias in LLMs are challenging tasks, requiring in depth understanding of the different phases of LLM training [42].

*2. Factually inaccurate information*

Another commonly discussed limitation of LLMs is that they can provide factually inaccurate information. As mentioned earlier, LLMs are not database retrieval tools [43]. Their mapping of the training data is "compressed" in a lossy manner [44], meaning that the model cannot retrieve the data in its original form, and its output is always the result of on-the-spot text generation based on statistical inference[5] [45]. Even when the training data is factually correct, LLM-generated output can contain factually incorrect statements. Such behaviour is often referred to as *AI hallucination* [31]. Factually incorrect or made-up statements can be particularly problematic in fields such as law, medicine or history, but also elsewhere. Yet, the so-called hallucinations can also play a productive role in creative processes, where novel ideas not previously present in the training data are actually desired.

In physics contexts, and especially in the context of learning introductory physics, confabulated or untrue statements tend to manifest themselves somewhat differently, as retrieval of large amounts of factual information is not typically central to physics tasks. Mathematically incorrect statements or mistakes in logic, on the other hand, are more relevant for physics applications (and arguably harder to spot than, for example, a wrong value of an otherwise well-known physics constant). For example, researchers have shown that even basic operations, such as multiplication, can present a significant challenge to GPT-3 [46]. This is a direct consequence of the fact that LLMs are not actually performing calculations but are instead producing an answer using statistical inference from its learning data. Such "arithmetic confabulations" nicely illustrate the limitations of the LLM functioning. Even if the data contains many correct examples of numbers being multiplied and the result of the multiplication, the model cannot extract and implement a general rule for multiplying arbitrary pairs of numbers and on its own does not have the capacity to judge the correctness of the answer, nor can it relate critically to its own capability to multiply numbers. Indeed, OpenAI's state-of-the-art application, ChatGPT-4, as of summer 2023, still cannot reliably multiply four-digit numbers and regularly returns incorrect results when used in "default mode", which relies on the underlying LLM as the source of information for generating responses.

---

[5] This is also why LLM-generated text often goes undetected though plagiarism detection software, which looks for direct matches of a piece of text against large databases.

*3. Plugins*

In summer 2023, OpenAI introduced a beta feature called "plugin mode", in addition to the "default" mode, for subscription-paying users of ChatGPT-4. A plugin is an external module, which is typically tasked with operations that the LLM cannot perform adequately. Plugins, such as Wolfram represent an attempt at improving ChatGPT's mathematical abilities [47]. Other kinds of plugins have also become available for enhancing ChatGPT-4's performance[6]. However, for this approach to work well, the chatbot would need to reliably make autonomous decisions about when to outsource tasks to external modules. Achieving this remains a challenge for LLM and chatbot developers. However, when such a decision can confidently be made by the user, they can explicitly instruct the chatbot to outsource certain operations. For example, a user can enable the Wolfram plugin [47] and instruct the chatbot explicitly to use it in solving a computational task. This way, plugins can, when in the hands of experienced users, significantly enhance the usability of ChatGPT-4 for tasks that extend beyond what LLMs can reliably do on their own. Another example of the plugin approach to the enhancement of ChatGPT's computational abilities is the introduction of a beta feature, "Advanced data analysis mode" (previously known as the Code interpreter plugin), made available to subscription-paying users of ChatGPT in late summer 2023. It introduces a code-compiler within the chat window, allowing ChatGPT-4 to not only write, but also execute Python code. This feature significantly improves the chatbot's ability to perform calculations and other tasks that have proven to be challenging to LLMs without plugins, also referred to as ChatGPT-4's "default mode". While plugins are a relatively recent development on the chatbot landscape, in the future they will likely play an important role in how LLM-based tools are used in a wide range of contexts, including physics and physics education.

For generating examples in this paper, we used ChatGPT-4 in its "default mode", which does not make use of plugins. We did this for two main reasons. First, this paper focuses predominantly on tasks that do not involve computation. Second, we believe that having a basic understanding of LLMs functioning without plugins is necessary for the productive use of plugins down the line. However, a systematic and in-depth discussion of plugins falls beyond the scope of this paper.

*4. Problematic argumentation*

Another challenge for the use of LLMs in the domain of physics, is that they often produce incoherent or inconsistent chains of argumentation. For example, in introductory classical mechanics, the challenge rarely lies in being able to merely recite known conceptual and mathematical relations that describe physics models, or to perform calculations. In fact, the relations and fundamental principles are so few and consistently formulated in the existing texts that LLMs typically have no issue producing them without mistakes. The real challenge lies instead in the consistent application of basic principles to find ramifications of physics models and generate predictions about the physical world [48]. Recalling a solution to a problem is rarely the point of learning physics [49]. On the contrary, it is how one arrives at the solution that matters. This typically requires using both mathematical and physics concepts. Physics thus presents a particularly interesting context for exploring LLMs' abilities to produce coherent conceptual and mathematical argumentation.

Below, we provide an illustrative pair of examples demonstrating how ChatGPT-4 deals with a conceptual introductory physics question that does not involve any numerical calculation. The examples showcase that ChatGPT-4 is still quite unreliable at answering such questions.

---

[6] As of November 1, 2023, more than 1000 different plugins are available to users of ChatGPT-4.

> *If two bodies with different mass have the same kinetic energy, which one has the largest momentum?*

We regenerated the response in ChatGPT-4 eight times. We got the correct answer that the body with the larger mass has the larger momentum in four out of the eight responses. Below, we first present an example of a response with a correct conclusion, even though a more thorough explanation would be desirable, especially towards the last part of the response.

> *The kinetic energy (K) of a body is given by the formula:*
>
> *K = 1/2 * m * v^2*
>
> *where:*
> *m is the mass of the body, and*
> *v is the velocity of the body.*
>
> *The momentum (p) of a body is given by the formula:*
>
> *p = m * v*
>
> *where again:*
> *m is the mass of the body, and*
> *v is the velocity of the body.*
>
> *If two bodies have the same kinetic energy but different masses, it means that the body with the smaller mass must be moving with a higher velocity to make up for the smaller mass in the kinetic energy formula.*
>
> *However, when you look at the momentum formula, it becomes clear that even though the smaller mass body is moving faster, it might not have more momentum than the larger mass body. This is because momentum is directly proportional to both mass and velocity.*
>
> *Therefore, for two bodies with different masses but the same kinetic energy, the body with the larger mass would have the larger momentum.*

We regenerated the response to this task through the ChatGPT-4 end-user interface by starting a new chat each time. For four out of eight responses, we got an incorrect conclusion that the object with the smaller mass has larger momentum. Below, we now also present one example of an incorrect answer:

> *Kinetic energy (KE) and momentum (p) are related to the mass (m) and velocity (v) of an object as follows:*
>
> *KE = 1/2 * m * v^2*
> *p = m * v*
>
> *If two bodies have the same kinetic energy, but different masses, we can analyze the situation using the above formulas.*
>
> *Let's call the two objects A and B, with mass mA and mB (mA > mB) and velocity vA and vB respectively.*
>
> *Since the kinetic energy of both bodies is equal:*
>
> *1/2 * mA * vA^2 = 1/2 * mB * vB^2*
>
> *Solving for vA and vB from here:*
>
> *vA = sqrt((2 * KE) / mA)*

*vB = sqrt((2 * KE) / mB)*

*And since mA > mB, it follows that vA < vB.*

*The momentum of the two bodies would be:*

*pA = mA * vA*
*pB = mB * vB*

*Given vA < vB and mA > mB, we can't directly determine which momentum is greater without knowing the specific values of mA and mB. However, because momentum is linearly proportional to velocity while kinetic energy is proportional to the square of velocity, the object with the lower mass (mB) will have a higher increase in velocity than the decrease caused by its lower mass. Therefore, the object with less mass (and higher velocity) will have a greater momentum when they have the same kinetic energy.*

For this particular question, ChatGPT-4 returned a combination of mostly correct argumentation and a correct final response only for half of the attempts. Importantly, our simple investigation does not provide us with a definitive measure of the rate of correctness of ChatGPT-4's responses to physics questions but rather allows us to illustrate our own and others' findings that the chatbot is unreliable and inconsistent even on basic conceptual problems. This unreliability and inconsistency in producing well-argued and correct responses, combined with what looks like an authoritative, plausible-sounding and confident tone, presents an important limitation for the use of ChatGPT-4 in physics education [17,24,50], especially if students wish to use it as a tutor. We will further discuss some possible ways of using LLMs in education in section V.

### D. LLMs vs. humans

From the brief overview of LLMs' functioning provided in this paper, we can already see at least two important ways of how their way of working fundamentally differs from that of humans.
First, humans need far less linguistic input information to be able to produce text which is comparable to even the most state-of-the-art LLM. To produce coherent and plausible-sounding output, LLMs need to be trained on truly huge datasets, which a single person could not possibly consume in a lifetime or even multiple lifetimes.
Second, in addition to written language, human discourse is influenced by other inputs. This is because people inhabit a world which is itself a source of non-linguistic information (e.g., visual, auditory, haptic). In addition, human learning is heavily influenced by implicit and explicit social rules and norms. In contrast to humans, LLMs lack non-linguistic real-world experience. This is one of the reasons an LLM's output can sometimes be incorrect in ways that feel uncanny to a human reader. We have, for example, noticed that its answers to problems on the topic of kinematics or dynamics (with a strong everyday coupling) sometimes contain mistakes that a human would be very unlikely to make, such as that an object thrown vertically continues travelling upwards even after it has reached the highest point on its trajectory [17].
On the other hand, LLMs can also produce incorrect statements that mirror learners' difficulties [17,51]. Talanquer even suggested that this correspondence may reveal some underlying similarities in the process of the emergence of students' and LLMs' difficulties [51].
Despite our understanding of the differences between how LLMs and humans function, anthropomorphising is relatively widespread in the discourse about LLMs, even in the research literature on the topic. Humans, including researchers, often explicitly or implicitly attribute to LLMs abilities like "thinking", "reasoning", "understanding", and "knowing" [30] and, as we mentioned before, "hallucinating"[7] [31]. This is understandable since state-of-the-art LLM-based chatbots have been intentionally designed and fine-tuned to be human-like in their manifest behaviour.

There are, however, important pitfalls of anthropomorphising LLMs. Here, we only briefly touch on two general ways it can be problematic: technical and affective. From the technical point of view, adherence to the interaction practices used in social situations between people can represent a barrier to using LLMs efficiently [53]. To make the best possible use of LLMs, they need to be prompted in specific ways by providing several categories of information related to the particular domain of the question at hand and other specifications that we usually would take for granted in a conversation with a person. This topic will be further explored in section IV, dealing with prompting techniques. Emotion-wise, important considerations need to be done when we allow ourselves or our students to spend time interacting with an LLM-based chatbot, especially if this involves long and intensive periods of interaction. One such consideration is how interacting with chatbots makes us feel and how it impacts our self-perception as learners and/or experts [54]. This topic will be further discussed in section V.

Understanding the differences between humans and LLMs in terms of their functioning represents an important resource for using LLMs as tools. To make the best use of LLMs and have realistic expectations of what they can offer, we find it useful to keep in mind and repeatedly remind ourselves how LLMs differ from humans, even if, and especially because, they are designed to convince us they are human (namely, by emulating human discourse). This does not mean that all our intuitions about interpersonal communication are useless for interacting with LLMs, but rather that we should be prepared to be selective and critical of the use of these intuitions when interacting with LLMs.

## IV. IMPROVING THE QUALITY OF LLM RESPONSES

### A. Introduction to prompt engineering

A prompt is a natural language instruction given to an LLM to facilitate its generation of an output. Prompting can thus be seen as a form of programming used to instruct an LLM [55]. When dealing with a chatbot like ChatGPT specifically, a user prompt is a query that the user writes for the chatbot using the chat interface. How a prompt is formulated highly impacts the output of the model [56]. Therefore, learning to formulate appropriate prompts is crucial for the productive use of LLMs and LLM-based chatbots in any context, including physics.

The craft of formulating effective prompts is often referred to as *prompt engineering*. Prompt engineering is a highly empirical field. It is not uncommon to see an LLM fail a given task but significantly improve its performance once given a properly refined prompt [57]. So, it has been argued that an LLM's failure is not necessarily evidence of the lack of an LLM's capacity to solve similar problems. It could just mean that we have not yet learned how to formulate the prompt in a good way [58].

Designing an effective prompt is a non-trivial task, requiring at least some understanding of how LLMs work. Zamfirescu-Pereira et al. [53] observed how end users with zero or very limited familiarity with LLMs struggle with generating prompts, evaluating prompt effectiveness, and explaining the effects their prompts had on the LLM's outputs. Without any in-depth understanding of LLM mechanics or dedicated instruction on prompt engineering, the participants in the study tended to approach the formulation of inputs as if they were dealing with a human. The authors [53] suggest that social experiences and expectations influence the research subjects' attempts at prompting in a way that reduces the quality of the obtained LLM responses. This supports our argument that understanding LLMs functioning is useful if we are to make good use of them. As discussed above, an important point of departure for prompt engineering is the understanding that we are not dealing with a human. Furthermore, it is important to understand LLMs' inherent strengths and weaknesses. For example, LLMs are particularly well suited for tasks such as text summarization and paraphrasing but much worse at algebraic tasks and planning solutions to multiple-step problems, even those often perceived as trivial for humans [59]. With just this information in mind, we can already expect that solving physics problems will likely present a challenge for LLMs. In our case, this expectation stems not only from consulting others' research but also from first-hand experience interacting with LLMs and asking them conceptual physical questions (see examples presented in this paper).

---

[7] The term hallucination has been suggested to be misleading, because it implies that there is a perceptual component to LLMs' acting. Alternatives to the term such as simply "making things up" or even "bullshitting" [52] have been argued to be more representative of the mechanism of the production of these statements.

However, we will show that with some relatively simple prompt engineering techniques, one can significantly improve the performance of ChatGPT-4 on such physics tasks. By showing some illustrative examples, we aim to demonstrate how the understanding of LLMs' functioning and applying this knowledge through prompt engineering can improve the correctness and reliability of ChatGPT-4's responses on introductory physics tasks.

Here, we would also like to issue a word of caution. Due to the probabilistic mechanics of LLMs, it is unlikely that using the same prompts will produce the same outputs every time. The exact reproducibility of the examples shown in this paper should not be taken as the only criterion for the validity of our arguments. The tool we have used to generate the responses is still receiving updates, and its performance on any particular task may have improved since we generated the responses. Our aim is not to provide definitive prompts to help solve concrete examples of tasks but to raise awareness within the PER community of how LLMs work and demonstrate through illustrative examples how this knowledge can be put to use through different prompting approaches. We expect that this knowledge will be useful for some time to come, regardless of how well specific iterations of LLM-based tools perform on the concrete examples selected for this publication.

### B. Providing the context

In Fig. 1, we have already illustrated how the bare-bones GPT model acts as an advanced "auto-complete" tool, which continues the started sentence in a way that is statistically most likely, based on the regularities in the training data, and, importantly, the specific text given to it in the prompt. The prompt serves as a point of departure from which the LLM extrapolates and generates further text based on its pre-trained neural network. What makes GPT and similar modern LLMs powerful is their ability to direct and narrow their "attention" to a subspace of possible meanings. Fig. 1 demonstrates how GPT generates domain-specific explanations of the term force, not confounding the meanings of "force" across the domains of physics and law.

Taking GPT's context sensitivity and context-responsiveness into account is essential when crafting appropriate prompts. If we want to improve the model's performance in a given domain, we need to prompt it in ways that will elicit the relevant context for the response. Here, we provide an illustrative example of the impact of providing such context.

Taking inspiration from a video clip of a Nascar race ( [60], https://youtu.be/sOY9p5gFa5Q), we asked ChatGPT-4 to explain how a racer could gain an advantage in the last lap of the race by grinding the car against the outer fence of the racetrack. The trick used by the driver can be explained at a basic level with the application of Newton's second law in the context of circular motion. The driver took advantage of the outer fence to provide an additional force pointing towards the centre of the turn, allowing him to make the turn at a much higher speed than the other drivers, who were relying only on frictional forces between the tires and the road to keep them "in the turn".

We first prompted ChatGPT-4 with the following question and got the response that follows:

> ***A Nascar racer won the race in the last lap by grinding against the outer fence of the racetrack. Why did this trick work?***
>
> *Your question implies a situation where a NASCAR racer won a race by driving his car against the outer fence or wall of the track on the final lap. This scenario is interesting, but in reality, deliberately "grinding" against the outer fence of a racetrack isn't typically a successful strategy, due to several reasons:*
>
> *1. Damage to the Car: Scraping against the fence can cause damage to the vehicle, affecting its aerodynamics, tire condition, and overall performance. Significant damage could even lead to a tire puncture or mechanical failure.*
>
> *2. Loss of Speed: Friction between the car and the wall would typically cause a reduction in speed. In general, racers try to maintain a clean line to maximize their speed.*
>
> *However, there are situations where close wall proximity might be advantageous:*
>
> *1. Late Apexing: If the driver took a "late apex" line, where they approach the corner wide, hit the apex late, and then exit the corner wide, it can sometimes allow for a faster exit speed, which could be beneficial on the final lap if the straightaway follows. This technique may give the appearance of 'grinding' against the outer fence.*
>
> *2. Drafting and Slingshot Maneuver: If the move against the outer wall was to escape from the turbulence created by the car in front, or to line up for a "slingshot" pass down the following straight, it could possibly make sense. However, this still typically involves avoiding contact with the wall.*
>
> *In NASCAR races, it is common to see cars running very close to the wall, especially at high-speed tracks. This is because the banking (the incline of the track from the inside to the outside) can provide additional grip, and the higher line can allow drivers to maintain higher speeds. But actually making contact with the wall is generally detrimental to performance.*
>
> *Lastly, it's worth considering the possibility that the story might be somewhat embellished or misinterpreted. Motorsport, like any other sport, sometimes has its share of "fishing stories" where the reality might be a bit less dramatic than the tale being told.*

The chatbot did not provide physics argumentation for the success of the driver's trick. In fact, it did not really answer the question. It did demonstrate some knowledge of the Nascar racing domain (although it is hard for us to judge the quality or reasonableness of the response in this context without being experts on the topic). It used technical terms and explained what appear to be typical approaches and tactics used by drivers in certain situations. Throughout the response, it maintained what looks like a sceptical tone, and in the conclusion even suggested that the story might have been manipulated or exaggerated.
If we are equipped with some basic understanding of LLM functioning, it is unsurprising that the response is firmly rooted in Nascar racing jargon. To the LLM, there is very little in the prompt that would cue an introductory-mechanics approach to answering the question. Other researchers have also reported that ChatGPT can have trouble answering simple physics problems when they are not contextualised explicitly as physics problems [22].

*1. Specifying the domain*

The first technique for improving the response is adjusting the prompt by specifying the domain on which the response should draw. Specifying the domain helps the LLM narrow down the context from which it draws inferences when generating the response. In our case, the aim is to get ChatGPT to not focus on Nascar as the primary domain of information but rather contextualise the answer in physics language. We therefore asked the chatbot to explain the situation from a perspective of forces in circular motion, resulting in the following refined prompt and response:

> **A Nascar racer won the race in the last lap by grinding against the outer fence of the racetrack. Why did this trick work? <u>Explain it from the perspective of forces in circular motion.</u>**
>
> *This maneuver appears counterintuitive because generally, you would want to minimize friction and contact with anything that could slow you down. However, in the context of circular motion and NASCAR racing, this can actually be a smart tactic due to the principles of physics.*
>
> *There are a few factors that could explain why grinding against the outer fence during the last lap might help a racer win:*
>
> *1. Centripetal Force: When a car is moving in a circular path, like on a racetrack, there needs to be a force that is directed towards the center of the circle to keep the car moving along that circular path. This is known as centripetal force. In the case of a NASCAR race, the friction between the tires and the track usually provides this force. However, by grinding against the fence, the driver may have been able to use the additional friction between the car and the fence to create an additional centripetal force. This could potentially allow the car to maintain a higher speed without skidding outwards due to centrifugal force (the apparent force that draws a rotating body away from the center of rotation).*
>
> *2. Banking of the Turns: NASCAR tracks are typically banked to help the cars make the turns at high speeds. The banking allows the normal force exerted by the track (which is perpendicular to the surface of the track) to provide a component of the centripetal force required for circular motion. If the racer is at the top of the banking (near the fence), the angle of banking is steeper, and hence the component of the normal force providing the centripetal force is larger. Therefore, he can maintain a higher speed.*
>
> *3. Drafting and Aerodynamics: NASCAR drivers often use a technique known as "drafting" where they align their cars with another car in front to reduce wind resistance. The leading car blocks the air for the following car, reducing the following car's aerodynamic drag and allowing it to save energy and increase speed. If the driver was following closely behind another car while grinding the fence, it could have given them an aerodynamic advantage.*
>
> *This is a simplified explanation and the real situation can be more complex, taking into account factors such as the condition of the tires, the car's handling characteristics, the power output of the engine, and the driver's skill and experience. But from the perspective of forces in circular motion, these could be potential reasons why grinding against the fence could have given the racer an advantage.*

The response is now more clearly focused on physics argumentation, providing a few possible physics explanations for why the "grinding" trick may have worked. Note that in this case, the response actually addresses the question instead of arguing that the car might have only appeared to be grinding against the fence. However, only the first bullet addresses the phenomenon directly and suggests incorrectly that friction between the wall and the car contributes to the "centripetal" force, as well as mentions centrifugal force, which is not required to explain the phenomenon. We can say that the new prompt provides a step in the right direction, but the response is still not particularly high quality from a physics perspective.

It is worth noting that the appropriate domain can be cued explicitly, as demonstrated above, or implicitly, by formulating the problem in terminology closely associated with a particular domain. For example, end-of-chapter problems in textbooks often contain phrases such as "consider a point mass" or "determine the coefficient of friction", which typically cue the LLM to answer the question using language from the physics disciplinary domain without needing an explicit prompt to do so. However, when faced with everyday situations, such reformulation of problems into physics language already involves a fair amount of physics modelling.

## 2. Specifying how to act

Another well-documented prompting technique [61] is to ask the chatbot to behave like an expert in a given field of interest. This tactic can be seen as a variant of the "specifying-the-context" tactic. The previous prompt can be further tweaked by specifying how the chatbot should act. In our case, to get a better physics explanation, we asked the chatbot to act like a physics teacher. The updated prompt was:

> **A Nascar racer won the race in the last lap by grinding against the outer fence of the racetrack. Why did this trick work? <u>Explain it like a physics teacher would from the perspective of forces in circular motion.</u>**
>
> *Sure, I'd be happy to explain this from a physics perspective!*
>
> *When a car is moving in a circular path, as it does around a NASCAR track, it experiences a force known as centripetal force. This force is directed towards the center of the path, and it's required to keep the car moving in a circular trajectory. Without this force, the car would simply continue in a straight line according to Newton's first law of motion (the law of inertia).*
>
> *The source of the centripetal force in this case comes from the friction between the tires of the car and the surface of the racetrack. However, as the car speeds up, more and more friction is required to supply the necessary centripetal force. Beyond a certain speed, the tires can't supply enough static friction and they start to slide - a situation known as understeer or oversteer. In either case, control over the vehicle is lost.*
>
> *Now, when the racer moved his car against the outer fence, what he essentially did was introduce an additional force to help supply the required centripetal force. This force came from the contact of the car with the fence. By pushing against the fence, the car could maintain its circular path at a higher speed than if it relied on tire-road friction alone. This allowed the driver to maintain a higher speed through the turn without losing control, enabling him to overtake competitors.*
>
> *However, it's important to note that this isn't an optimal way of completing a race. Grinding against the fence would cause significant damage to the car and could even lead to a dangerous situation. It's a tactic of desperation, and it only worked here because it was the final lap - the car just needed to hold together for a little longer. But, from a physics point of view, it makes sense - by utilizing the fence as an additional source of centripetal force, the car could take the curve at a higher speed than would be possible relying solely on the friction between the tires and the track.*

This time, the answer is correct and quite well-formulated from a physics and physics teaching perspective. At the same time, it still contains a healthy amount of scepticism towards the reasonableness and safety of the described racing tactics. We can say that it appropriately grounded its response in both physics and racing contexts.

*3. Comments on providing the context*

LLMs' context sensitivity and responsiveness can be seen as both a strength and a weakness. The weakness aspect becomes apparent when the prompt contains superfluous information. Shi et al. [62] show how superfluous details can distract an LLM and degrade the quality of its responses. It can be difficult for an LLM to determine which information is crucial for the task. This can also be observed in physics examples. We experimented with providing ChatGPT-4 with problems concerning an incline but instead of using a generic "mass" for the sliding object sliding down the incline, we introduced everyday objects, such as a shoe or a *Cornu aspersum* (the Latin name for the common garden snail). We observed that the chatbot's responses were much less "to the point" in these cases (from a physics point of view) and contained many non-physics references (zoological references in the case of the snail). This resembles how the Nascar context in the above-presented example significantly influenced the content of the response.

In another example, we saw that specifying in the problem text that the setup was a "modified Atwood machine" (where one of the blocks is placed on a horizontal frictionless surface, and the other hangs vertically)[8] resulted in ChatGPT-4 producing a mathematical expression corresponding to an ordinary Atwood machine (where both blocks hang vertically). Given how LLMs function, this is not particularly surprising. Statistically, it is more likely that a problem containing the phrase "Atwood machine" would have in its solution the expression for an ordinary Atwood machine, even if the problem actually deals with a "modified" Atwood machine.

On the other hand, with some basic knowledge of how LLMs work and the use of simple prompt engineering techniques, context sensitivity and responsiveness can be productively harnessed to generate output which is well-focused in terms of both content (by specifying the context)[9] and style (by specifying how to act). We have also illustrated in this subsection that these prompting techniques can, besides generating the output in more contextually appropriate language, also improve the correctness of the output content from a physics perspective.

It is once again important to be aware that LLMs are not human and thus often require different approaches to prompting. While context is often made available to humans in different ways, including non-linguistic cues like being physically present in a physics class or inhabiting specific social roles, such as a student or a physics teacher, LLMs often need to be prompted explicitly to draw on specific contexts or to imitate specific types of discourse. We need to remind ourselves that prompts that would be seen as unnecessary or even rude when interacting with people, such as high repetitiveness [53], may present a powerful prompting technique when interacting with LLMs.

In addition to using prompts to direct the LLM to draw on specific domains of knowledge from its training data, they can sometimes also be used to provide the LLM with new information not included in the initial training. The extent of such new information can vary significantly. It can range from introducing an important fact that the LLM should consider when formulating the response to providing it with a whole scientific article or even a book (if the LLM's context window is large enough). A common term for using the prompt to "teach" an LLM new information is called In-context Learning (ICL).

In the following section, we will briefly explain the history and early uses of ICL in the LLM research literature and how it relates to another important family of prompting approaches – *Chain-of-Thought* prompting.

---

[8] The problem was taken from [26 p.110].
[9] Another related technique to strenghten the grounding in a specific context is to write a prompt where essential information is repeated multiple times. Zamfirescu-Pereira et al. [53] consider this technique a useful but under-investigated technique in prompt engineering.

## C. The Chain-of-Thought approaches

Physics problem solving (real-world or end-of-textbook) is typically a multi-step process, often involving running mental models of physical phenomena, translating between the physical world, mental models and mathematical formalisms [48]. One typical goal of physics instruction is for students to learn to reason systematically and build robust arguments from basic physics principles. By having some understanding of how LLMs work, we can see that their functioning is almost the opposite of what we want our students to do. While LLMs can give the illusion of reasoning by imitating disciplinary discourse, they do not "think" before they "talk" [30]. It is, in fact, quite remarkable that LLMs can produce what on the outside appears like logically coherent arguments [63]. In computational linguistics, this ability is often referred to as the ability to "reason" despite this obviously being a misnomer[10]. In computational linguistics, anthropomorphising terms like "reasoning" and "thought" are often used to refer to the abilities of LLMs, which we would typically ascribe to people displaying similar behaviours. In the remainder of this section, we also use the terms "reasoning" and "thought" in a way that reflects their use in the existing literature on LLMs. However, we wish to stress once again that the process behind this apparent reasoning is very dissimilar to human reasoning and would be better referred to as "argumentation" to avoid the often-misleading anthropomorphising of LLMs.

Chain-of-Thought (CoT) approaches to prompting are a family of prompting strategies to improve the performance of LLMs on complex reasoning tasks, such as arithmetic reasoning, common sense reasoning, and symbolic reasoning [57,64]. CoT prompting engages the LLM in step-by-step "reasoning" and has been demonstrated to lead to better output than when an LLM omits such intermediate steps in its response [64].

The possibility to "think out loud" is crucial for LLMs' performance since they cannot "reason" other than through generating text. By literally spelling out the reasoning, every step of the process of solving a task is effectively appended to the prompt, and the "auto-complete instinct" of the LLM allows it to build a chain of argumentation, which can exhibit continuity and logical coherence[11]. As we will demonstrate below, CoT prompting holds the potential to improve LLMs' likelihood of correct "reasoning" and quality of the "reasoning" itself in the domain of physics. First, we provide some background that will allow for a better understanding of how and why CoT prompting is such a powerful approach to improving the quality of LLM output. We begin by reviewing the evolution of the CoT approach from the so-called "In-context Learning" approaches to prompting.

### 1. From In-Context Learning to the Chain-of-Thought strategies

Historically, the term In-context Learning (ICL) was used to describe a family of approaches aimed at improving the performance of LLMs. ICL takes advantage of the capacity of LLMs to learn from the prompt itself instead of just its training data [4]. ICL presents an important and less resource-intensive approach to teaching LLMs, as it does not require revisiting the resource-expensive pre-training procedure. One important limitation is the length of the prompt, which can only be as long as the particular LLM's context window (roughly analogous to the LLM's working memory). In the case of the more recent GPT versions, the maximum context length is quite limited: GPT-3.5 allows for about 3000 words, and GPT-4 allows for about 6000 [67]. New developments in LLM design, however, may soon allow for a significant expansion of prompt length, which also expands the possibilities for ICL. One possible use case in physics would be feeding the LLM a physics textbook to facilitate successful end-of-chapter problem-solving using a particular approach explained in that book. However, this has yet to be tested in practice. For a comprehensive review of ICL approaches, see [68]. A common approach to ICL is to begin the prompt by providing one or more examples of tasks accompanied by solutions, to create a demonstration context, and then providing the task to be answered by the LLM. In this way, the LLM can learn from analogy directly from the prompt by means of imitating it. Fig. 2 shows an example of building the prompt with a series of *input/output pairs*.

---

[10] Granted, the computational linguistic community is likely less concerned with the epistemological considerations of physics learning than the PER community.

[11] Cognitive linguists have compared this step-by-step approach to solving problems to "thinking slow", by analogy to the dual-process theories of human cognition [65,66].

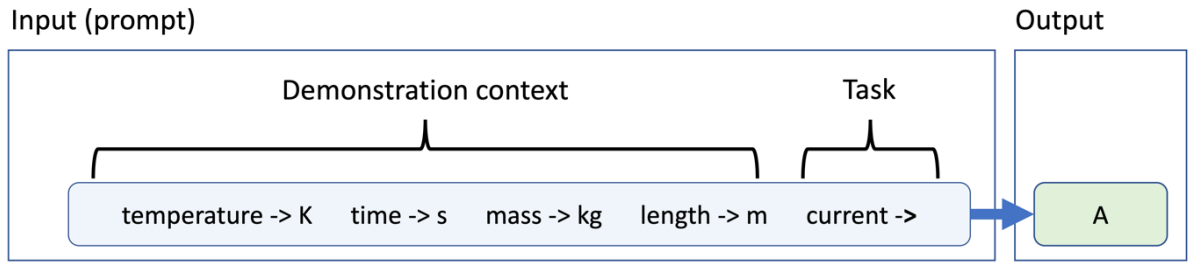

**Figure 2**. Illustration of an In-context Learning (ICL) prompt. The prompt begins by providing input/output pairs as the demonstration context (assigning SI-unit symbols to physical quantities). The demonstration context is followed by a task, left open for the LLM to complete. The generated output is shown on the right.

ICL approaches can work even with few examples and without detailed task specifications. Once again, this approach leverages the LLM's capability to produce a plausible continuation of the sequence presented in the prompt. This type of short example-based ICL works well with simple tasks. However, providing examples for more complex tasks becomes more challenging.

## 2. Chain-of-Thought prompting

More recently developed LLMs like GPT-3.5 and GPT-4 present emergent "reasoning" abilities, an unforeseen side effect of the higher number of learning parameters they are trained on [63]. This means that they are capable of producing arguments that resemble human reasoning on a variety of tasks. This skill is crucial for introducing LLMs into fields such as physics, which often require advanced argumentation and reasoning capabilities. It turns out that the ability of LLMs to exhibit reasoning can be exploited by providing the LLM with more articulated demonstration examples as part of ICL. Because of the complexity of the tasks, the provided pattern can no longer be simple input/output pairs. The examples need to showcase desired forms of reasoning. This brings us to the Chain-of-Thought (CoT) family of approaches to prompting. CoT strategies have been developed building on the existing idea of ICL and have been demonstrated to improve the performance of LLMs [64]. In addition to providing input/output pairs as a demonstration context, CoT prompting incorporates a series of intermediate reasoning steps into the prompt. A complex task can often be approached by decomposing it into sub-tasks and then solving these smaller parts one by one. This is also true for many physics problems.
In the following subsections, we elaborate on the use of CoT prompting. CoT prompting can be integrated into ICL approaches by providing the LLM with a few examples of how to reason (*Few-Shot CoT* prompting) or by explicitly instructing it to reason step-by-step without providing any concrete example of a solution in the prompt (*Zero-Shot CoT* prompting). Moreover, recent and ongoing research is providing further prompting strategies, expanding the CoT family of approaches (*Self-Consistency* [69], *Tree-of-Thought* [70,71]), including approaches involving a recursive dialogue between the end user and the LLM [72] or even between LLMs [73].
While these approaches have shown great promise in improving the quality of LLM output, it is important to recall that the probabilistic nature of LLMs means that these improvements are typically achieved statistically over a large number of produced outputs [64], and individual outputs may vary in quality.

### 3. Few-Shot CoT prompting

Few-Shot Learning is the direct evolution of ICL for more complex reasoning tasks, which incorporates the CoT reasoning steps into the input/output demonstration provided as learning context in the initial prompt. This approach can be successful even with just one example: in this case, we refer to the strategy as One-Shot Learning. Fig. 3 shows an example of its functioning.

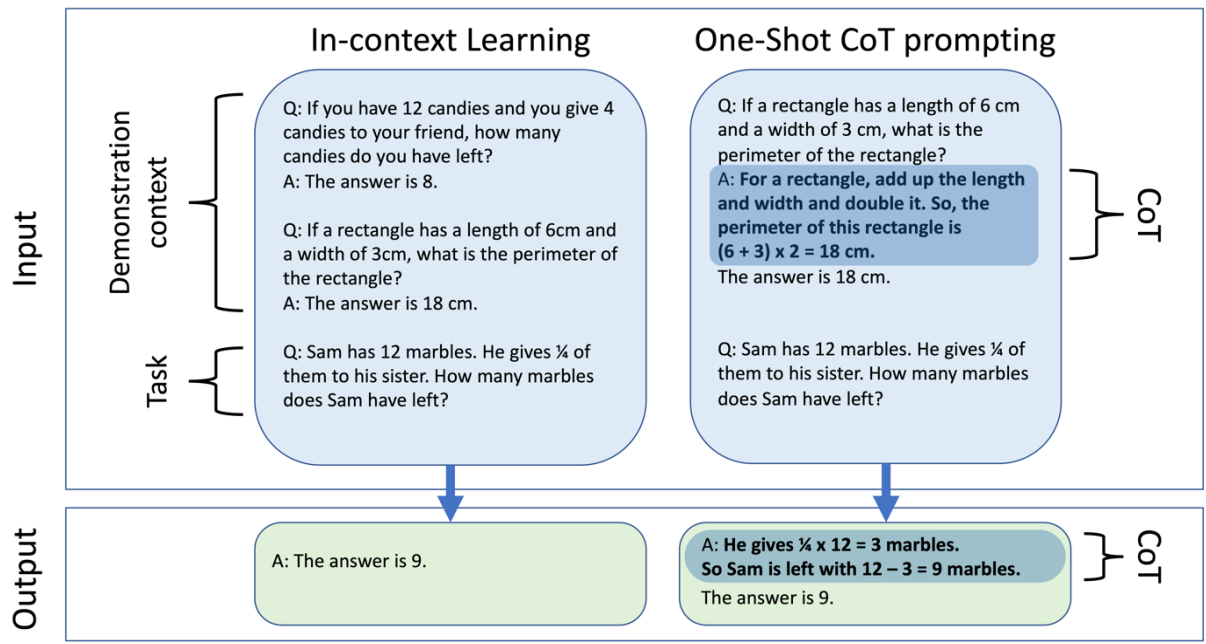

**Figure 3**. Example of a Few-Shot-ICL prompt (left) vs. One-shot-CoT prompt (right). Image adapted from [74].

In this way, the reasoning proposed in the prompt becomes crucial since the model will base its response on it. However, providing isomorphic examples with correct and well-structured solutions is not necessarily a straightforward task. In physics, being able to identify the domain and structure of a problem and providing a similar problem with a good solution as a learning demonstration for the LLM requires a good knowledge and understanding of the topic. This is likely to represent a major difficulty, especially for students in their early phase of learning. Consequently, this strategy is not easily applicable for novices or in fields where the users lack previous experience.

### 4. Zero-Shot Chain-of-Thought prompting

Zero-Shot CoT is another strategy belonging to the family of CoT approaches, which prompts the LLM with an explicit request to write out its reasoning. This method is more straightforward to implement than the Few- or One-Shot CoT approach since it does not require examples of reasoning as part of the prompt. The core idea is instead to stimulate the LLM to engage in step-by-step "reasoning" by simply asking it to do it, as illustrated in Fig. 4. This approach leverages the capability of the LLM to imitate the human form of step-by-step reasoning, which it had encountered in its training data. Encoded patterns of step-by-step reasoning in the pre-training data thus serve as a general resource that can be employed in concrete tasks where explicit reasoning is prompted for. In this sense, CoT prompting essentially provides further context to the LLM. In this case, the context is related to the general form of the response. The addition of a CoT prompt effectively makes the LLM take a different path through its training data, one that more closely resembles careful deliberation in humans.

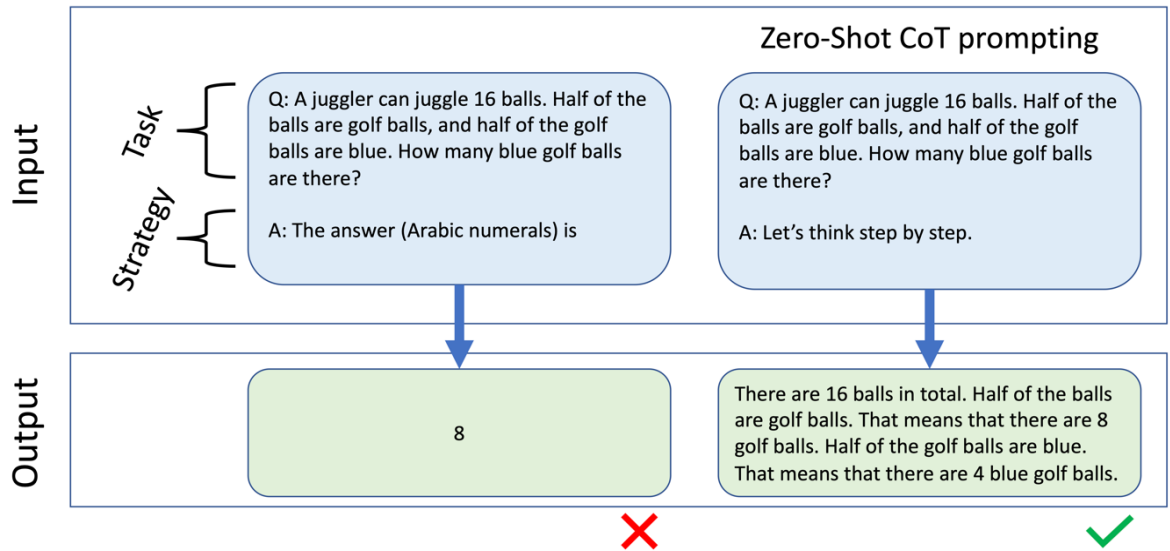

**Figure 4**. Example of a Zero-Shot prompt without CoT (left) and with CoT (right). Image adapted from [57].

This approach turns out to be successful for its simplicity and high effectiveness [57]. In fact, using it does not require any knowledge of the topic nor particular prompt engineering strategies. Research shows that Zero-Shot CoT prompting tends to improve the likeliness of correct responses and the quality of reasoning compared to non-CoT prompts [64].

We have tested the Zero-Shot CoT approach on several tasks, including the following free-response task, already presented at the beginning of the paper:

> *If two bodies with different mass have the same kinetic energy, which one has the largest momentum?*

With the original form of this prompt, ChatGPT-4 answered incorrectly four out of eight times upon regenerating the response. Moreover, even the responses containing the correct final answer mostly contained incomplete or partly faulty argumentation. However, upon applying a Zero-Shot CoT prompt, the results improved significantly. Kojima et al. [57] proposed the "Let's think step by step" line as a template to cue CoT reasoning. Other possibilities, such as "Explain your reasoning", have also been explored [57]. We formulated our Zero-Shot CoT query as follows:

> *If two bodies with different mass have the same kinetic energy, which one has the largest momentum?* <u>*Provide your reasoning first and only then provide the answer.*</u>

The rate of the correctness of responses increased to seven out of eight. The provided reasoning also became more detailed and logically well-structured. Here, we offer a representative example of a response to the prompt with a Zero-Shot CoT suffix underlined above:

> *Sure, let's break it down.*
>
> *The kinetic energy (K.E) of a body is given by the formula:*

> K.E = 1/2 mv^2
>
> where m is mass and v is velocity.
>
> The momentum (p) of a body is given by the formula:
>
> p = mv
>
> where m is mass and v is velocity.
>
> Given that two bodies with different masses have the same kinetic energy, this implies that the body with the smaller mass must have a larger velocity to achieve the same kinetic energy.
>
> Looking at the momentum formula, it's clear that momentum is directly proportional to both mass and velocity. Given two bodies with the same kinetic energy but different masses, the body with the smaller mass must have a greater velocity to compensate and achieve the same kinetic energy.
>
> Given the relationship between mass, velocity, and momentum, the body with the larger velocity (i.e., the one with the smaller mass) would have a larger momentum if the increase in velocity is enough to offset the decrease in mass.
>
> So, let's use some algebra to derive a direct relationship between the kinetic energy and momentum. From the formula of kinetic energy:
>
> v^2 = 2K.E/m
>
> Substituting v from the formula of momentum:
>
> p^2 = 2mK.E
>
> Therefore, momentum squared is directly proportional to the mass. This means that the object with the greater mass, assuming they both have the same kinetic energy, has the greater momentum.
>
> To summarize, the body with the larger mass will have the larger momentum if two bodies with different masses have the same kinetic energy.

Below, we provide a pair of examples that demonstrate an important feature and advantage of CoT prompting. CoT prompting shapes the LLM's output by influencing its reasoning path, making it more likely to follow a consistent chain of reasoning from beginning to end. We first present an example of a prompt without a CoT suffix and analyse how the answer went awry.

> **You hold a block on a horizontal, frictionless surface. It is connected by a string that passes over a pulley to a vertically hanging block. What is the magnitude of the acceleration of the hanging object after you release the block on the horizontal surface? (a) Less than g (b) More than g (c) Equal to g (d) Any of the answers a–c, depending on the masses of the blocks.**
>
> (d) Any of the answers a–c, depending on the masses of the blocks.
>
> The acceleration of the hanging object depends on the ratio of the masses of the two blocks. To understand this, we can analyze the situation using Newton's second law (F = ma) and considering the tension in the string and the gravitational force acting on the hanging block.
>
> Let m1 be the mass of the block on the horizontal surface, and m2 be the mass of the hanging block. When the blocks are released, the tension in the string (T) and the gravitational force (m2 * g) act on the hanging block. We can set up two equations based on Newton's second law:

> *For the horizontal block (m1):*
> *T = m1 * a*
>
> *For the hanging block (m2):*
> *m2 * g - T = m2 * a*
>
> *We can solve these equations simultaneously to find the acceleration (a):*
>
> *m2 * g - m1 * a = m2 * a*
> *a = (m2 * g) / (m1 + m2)*
>
> *From this equation, we can see that the acceleration of the hanging block depends on the masses of both blocks:*
>
> *(a) If m1 > m2, then a < g.*
> *(b) If m1 < m2, then a > g.*
> *(c) If m1 = m2, then a = g.*
>
> *So, the correct answer is (d) Any of the answers a–c, depending on the masses of the blocks.*

We can see that ChatGPT-4 first answers the question by selecting an answer (wrong answer), then proceeds to give a fairly well-argued analysis of the situation, including the derivation of the correct algebraic expression for the acceleration of the system. However, its initial selection of the answer (d) influences the final steps of the analysis, where it determines the values of acceleration for different values of the two masses. It appears as if the chatbot is justifying the initial answer (d) using this analysis, even if that means making, what are from a physics perspective, unjustified conclusions. We have seen this pattern reoccur also upon repeated generation of the response. If we understand the rudimentary principles of LLMs' operation, this behaviour should not surprise us. By generating each new word, the LLM effectively extends its prompt one word at a time. The generated words get added to the prompt and form a new context from which the subsequent word gets generated. Because the LLM is context-sensitive, it tends to build on the previously generated text. Because it does not have a mechanism for checking the correctness of the previously generated text, this can result in situations such as the one in the example above, where incorrect and correct statements and arguments are combined in what can appear weird or uncanny reasoning.

However, with the addition of a simple sentence at the end of the prompt, we can take advantage of the CoT approach:

> **You hold a block on a horizontal, frictionless surface. It is connected by a string that passes over a pulley to a vertically hanging block. What is the magnitude of the acceleration of the hanging object after you release the block on the horizontal surface? (a) Less than g (b) More than g (c) Equal to g (d) Any of the answers a–c, depending on the masses of the blocks. <u>Explain the reasoning first and only then provide the answer.</u>**
>
> *Let's consider the two blocks: block A is on the horizontal, frictionless surface, and block B is hanging vertically. The mass of block A is m_A, and the mass of block B is m_B. The gravitational acceleration is g.*
>
> *When you release the block on the horizontal surface, there are two forces acting on the system: the gravitational force acting on block B (F_gravity = m_B * g) and the tension in the string (T). Block A is only affected by the tension in the string, while block B is affected by both the tension in the string and the gravitational force.*
>
> *To analyze the system, we'll use Newton's second law of motion (F = m*a) for both blocks:*

*For block A (horizontal direction):*
*T = m_A \* a*

*For block B (vertical direction):*
*m_B \* g - T = m_B \* a*

*Substituting the expression for T from the first equation into the second equation:*
*m_B \* g - m_A \* a = m_B \* a*

*Now, we solve for the acceleration a:*
*a = (m_B \* g) / (m_A + m_B)*

*As you can see from the equation, the acceleration of the hanging object (block B) depends on the masses of both blocks (m_A and m_B). Since the denominator (m_A + m_B) is always larger than m_B, the acceleration a will always be less than g.*

*So, the magnitude of the acceleration of the hanging object after you release the block on the horizontal surface is:*

*(a) Less than g*

By avoiding making a fast conclusion at the beginning and effectively pushing the final judgement to the end of the response, the chatbot is able to arrive at the correct answer. It is important to note that regardless of which CoT-based prompting approach one might decide to use (Few-, One-, or Zero-Shot) the assessment of the correctness of the output remains in the hands of the end user.

## 5. Derivatives of the Chain-of-Thought prompting approach

The CoT family of approaches to prompting has received much attention in recent months. It has been found to be so successful that it is reasonable to expect that LLM-based applications will use it – and with high probability are already using it – "behind the scenes" to improve the performance of their products. For example, this can already be seen when using "prompt perfecting" tools, which utilise AI to improve user-generated prompts. We observed one such tool, a ChatGPT plug-in, append a Zero-Shot CoT prompt to our physics question when we asked it to improve our prompt. As of September 2023, ChatGPT-4 has increasingly been responding to different tasks in ways that resemble CoT without being explicitly prompted by the end user. We speculate that this may be the consequence of the results of research on prompt engineering being applied on a general level in the form of hidden pre-prompts, which are not visible to the end users[12] [75].

There are ongoing research and development efforts in the LLMs research fields that aim to develop even more advanced prompting strategies. Some are building on the CoT approach and can be seen as its derivatives. These derivative techniques aim to improve the LLMs' output quality by allowing the LLM to critically revisit its output. While an LLM cannot critically assess its own reasoning while generating it, it can do this with its past outputs. This can be done by having an LLM first generate some output and then include it in a new prompt asking for some form of critical revision.

In one strategy, referred to as *Self-Consistency* (SC) [69], the LLM is first prompted to generate several answers using CoT prompting and then asked to assess them before deciding on the final response. The idea behind this strategy is that the reasoning path underpinning the solution of a problem is not necessarily unique. Indeed, in physics, one can often approach a problem in different ways, producing different valid paths to an answer[13]. It could also happen that one path is better than the others because of its strength and consistency. The best response can be chosen by following a predetermined set of criteria or by deciding on the final answer by looking at the distribution of the answers generated through repeated prompting and selecting the most common one, assuming that correct answers are more likely to appear. This approach is more likely to work with simple problems, e.g. in arithmetic, and has yet to be tested in physics contexts and it is unclear if the SC's "democratic" approach can lead to more reliable and correct responses to physics tasks.

---

[12] However, because this is OpenAI's proprietary information, we are not able to verify our suspicion.

Another emerging strategy building on the CoT approach is known as *Tree-of-Thought* (ToT) [70,71]. It works similarly to SC, but instead of self-evaluating a reasoning path in its entirety, it evaluates every step (every "thought") in a longer argument. The main idea is that while generating text, the LLM can commit mistakes, because of its way of functioning (namely, not being able to stop and revise the output it is currently generating). However, when the model is given a text to evaluate, it is able to consider it as a whole and potentially revise it. In this way, the LLM can re-generate a new path (or several new paths) at each reasoning step, resulting in a "tree of thoughts", offering many possible reasoning chains.

## 6. Dialogue

A particularly promising approach to using the above discussed CoT-based prompting techniques in physics education is integrating them into human-LLM dialogue. There are two main advantages of doing this. The first advantage is that it does not require specialised technical programming skills and can be done by simply "chatting" to LLMs. Second, it can provide meaningful opportunities for eliciting reasoning about physics in students, or opportunities for physics teachers to hone their skills of giving helpful feedback to others' reasoning. In this way, some of our skills and intuitions around social interactions can be used productively for interacting with an LLM.

In previous research, one of the authors of this paper, together with a colleague, engaged in a Socratic-style dialogue with the first released version of ChatGPT in December 2022 [17] and found that pointing out inconsistencies in the chatbot's reasoning did not result in it improving its reasoning. This sort of behaviour of the chatbot the potential to cause feelings of frustration with the users, as has also been noted by Kortemeyer [19]. On the other hand, more recent models, such as ChatGPT-3.5 and -4, now project a "humbler" persona, one that is able to and often admits its mistakes or unclarities when they are pointed out or questioned. This lays a better ground for implementing dialogue-based prompting approaches [72] in physics contexts. The combination of these dialogue-based approaches with providing ample context and CoT prompting strikes us as an especially promising direction for future research efforts in PER.

To illustrate the possibilities of dialogue-based approaches to LLM prompting, we provide a concrete example of using a dialogue version of the SC approach to solve a physics task. The question is:

> ***There is a shoe that weighs 500 grams, resting on a table tilted at an angle of 30 degrees. What is the friction force between the table and the shoe?***

The problem describes a classical physics situation of an inclined plane and an object in static equilibrium on it. By employing Newton's laws for a body in static equilibrium, we can find the expression for the friction force, even without knowing the coefficient of friction. However, ChatGPT-4 frequently provided the following reasoning:

---

[13] It is important to remember though the possibility that one reaches the correct answer even when the reasoning path followed is incorrect.

> *To calculate the friction force between the table and the shoe, we first need to determine the normal force and then use the friction coefficient. However, you haven't provided the friction coefficient (usually denoted by μ) between the shoe and the table, so I will explain the general approach to solving the problem.*
>
> 1. *Calculate the gravitational force acting on the shoe:*
>    *F_gravity = m * g*
>    *where m is the mass of the shoe (0.5 kg, given that the shoe weighs 500 grams) and g is the acceleration due to gravity (approximately 9.81 m/s^2).*
> 2. *Calculate the normal force acting on the shoe:*
>    *F_normal = F_gravity * cos(angle)*
>    *where angle is the angle of the table (30 degrees).*
> 3. *Calculate the friction force:*
>    *F_friction = μ * F_normal*
>    *where μ is the friction coefficient.*
>
> *Without the friction coefficient, we cannot provide a specific value for the friction force. If you can provide the friction coefficient, I can help you complete the calculation.*

This reasoning path cannot be considered completely incorrect. However, it does not lead to an answer since it requires the knowledge of a parameter (the friction coefficient), which is not provided by the text. Our guess is that the formula with the normal force and the friction coefficient is statistically likely to appear together with "friction force" in the training data.

Our goal was to make it "consider" the possibility that there could be another way to approach the problem. We thus asked an additional question, which did not give any specific hints:

> ***Isn't there another way to do it?***
>
> *I apologize for any confusion. Yes, there is an alternate approach to find the friction force by considering the component of gravitational force acting parallel to the inclined surface of the table. Here's the step-by-step process:*
>
> 1. *Calculate the gravitational force acting on the shoe:*
>    *F_gravity = m * g*
>    *where m is the mass of the shoe (0.5 kg, given that the shoe weighs 500 grams) and g is the acceleration due to gravity (approximately 9.81 m/s^2).*
> 2. *Calculate the component of gravitational force acting parallel to the inclined surface:*
>    *F_parallel = F_gravity * sin(angle)*
>    *where angle is the angle of the table (30 degrees).*
> 3. *Since the shoe is at rest, the friction force acting on it must be equal and opposite to the component of gravitational force acting parallel to the inclined surface:*
>    *F_friction = F_parallel*
>
> *Now, let's do the calculations:*
> 1. *F_gravity = 0.5 kg * 9.81 m/s^2 = 4.905 N*
> 2. *F_parallel = 4.905 N * sin(30 degrees) = 4.905 N * 0.5 = 2.4525 N*
> 3. *F_friction = F_parallel = 2.4525 N*
>
> *So, the friction force between the table and the shoe is approximately 2.4525 N.*

The question *"Isn't there another way to do it?"* was enough to get the desired answer. This is a short and simple example of a dialogue with the chatbot, demonstrating the potential of dialogue-based SC-prompting for better performance.

However, we must also consider the possible limitations of using similar dialogue-based approaches. First, it is not self-evident that a chatbot would accept criticism or respond fruitfully to a follow-up question. Gregorcic and Pendrill [17] have reported that attempting to engage in Socratic dialogue with ChatGPT-3.5 in December 2022 was frustrating because the chatbot would not accept that it may have been wrong. Thus, it would have been harder to use it in an educational context for students. However, ChatGPT-4 is more likely to project a "humbler" attitude and accept the possibility that it may have gotten something wrong.

Still, the type of interaction shown in the above example requires the end user to have a specific mindset, which may not come naturally to physics learners. When a teacher provides a solution or answer to a physics task, the student typically does not ask, "Isn't there another way you could solve this?" Such a question would imply that the student has reason to doubt that the teacher has provided an exhaustive treatment of the problem. We believe that such an attitude is uncommon among physics learners. It is therefore crucial to consider what social norms and roles students might bring into their interactions with LLM-based chatbots and when these unspoken norms may actually hamper their productive use [53].

### 7. Comments on CoT and dialogue approaches to prompting

The step-by-step answers elicited by CoT prompting can be seen as particularly valuable in learning physics since they more explicitly delineate the path to a solution and consequently tend to arrive at better argued conclusions. Still, it is important to keep in mind that even though the quality of responses tends to increase when using CoT prompting, there is no guarantee that all parts of the chain of arguments, or the final solution, will always be correct from a physics perspective (as seen in examples). However, with the path to the solution being available for scrutiny, potential mistakes or unclear steps can be interrogated. This can be done by the LLM itself, through SC and ToT strategies, or by humans in dialogue with the LLM. In any case, having the reasoning process or the chains of "thought" available for scrutiny is helpful and necessary if we want to use LLMs productively in the process of learning and teaching physics. Furthermore, applying the understanding of LLMs and prompt- engineering skills in the context of dialogic interactions with LLMs should help us use them more productively.

## V. IMPLICATIONS FOR INSTRUCTION

In this section, we first summarise common topics in the existing literature on the instructional implications of LLMs. We only briefly overview general education literature and look more closely at physics-specific publications. We then propose how our own work presented in this paper relates to those considerations and what other implications we can see stemming from it.

The topic of artificial intelligence in education has been an established field of research. Publications in the field have explored the impact of AI on curriculum design and how it might impact the processes of teaching and learning [76]. However, since the public release of ChatGPT, there has been an escalation in the sense of urgency to investigate the possible implication of LLM-based tools for education [1]. In the months following the release of ChatGPT, scholars in different subfields of education research began reflecting on the opportunities and challenges arising from LLMs' application in education [77–81]. Some recurring ideas in this literature include the impact of LLMs on assessment practices on one side [52,82,83], as well as LLMs' potential to enable interactive and personalised learning assistance for students [84,85], and lesson planning and assessment assistance for teachers [84–88].

Researchers in PER point out that students having access to LLMs would likely require a revision of the typical tools for assessing and evaluating students' output. Yeadon et al. [23] demonstrated that ChatGPT-3.5 could achieve first-class performance on short-form physics essays in a university physics module on the history and philosophy of physics. They suggest that this holds serious potential to undermine the usefulness of at-home essay writing as an assessment tool. Kortemeyer's [19] findings suggest that ChatGPT-3.5 already holds the potential to significantly impact the use of programming in physics learning and assessment. While research has in some cases been clearly sceptical towards ChatGPT's usefulness for "cheating" [17], with further improvements in LLMs' development and the evolution of the end users' skills to prompt them effectively, we could soon see a more substantial impact on how physics teachers approach assessment and evaluation of physics knowledge and skills. Important questions include what physics-related skills students still need to master and how we should assess their mastery.

Regarding students' use of LLM-based chatbots, recent research points to some crucial issues regarding how students perceive their trustworthiness. Shoufan [89] explored senior engineering students' perceptions of ChatGPT-3.5, highlighting their admiration of its capabilities but also the need for a good background knowledge to assess its answers. As both Gregorcic and Pendrill [17] and dos Santos [18] pointed out, the linguistic proficiency of LLM-based tools can create an illusion of competence, especially when the end users lack an in-depth understanding of the produced content. Even more recently, Krupp et al. [90] have demonstrated that students who used ChatGPT as a tool in solving physics tasks often used its output uncritically, even in topics that they had significant experience with. In another study focusing on physics students' evaluation of ChatGPT performance, Dahlkemper et al. [91] asked students to analyse ChatGPT-3.5-generated answers to a set of questions of progressing difficulty in three areas of physics: rolling motion, waves, and fluid dynamics. They were asked to evaluate the answers in terms of their linguistic quality and scientific accuracy. Students were found to be less able to evaluate the scientific accuracy of answers on the topic with which they were less familiar – they had trouble distinguishing between a correct, expert-generated answer and an incorrect ChatGPT-generated answer. These research efforts suggest that a real drawback for learners might be an overreliance on AI-based tools, with a consequent risk of a reduction in critical thinking and problem-solving skills.

On the other hand, we can already see the first attempts of meaningfully integrating ChatGPT into physics learning activities to mitigate the challenges discussed above. Bitzenbauer [92] proposed using the chatbot in high school classrooms to foster students' critical thinking skills by asking students to use ChatGPT to generate a text about photons, evaluate its consistency and correctness, and revise it using additional sources. Gregorcic and Pendrill [17] have also suggested that ChatGPT's output could be used in the training of critical thinking in teacher education.

Against the background of existing literature on the educational implications of LLMs in physics and our own exploration of LLMs' functioning presented in this paper, we propose three possible roles[14] that LLMs could play in the process of physics teaching and learning: (1) LLMs as physics tutors, (2) LLMs as models of physics students, (3) LLMs as peers.

1. We have shown that without careful and deliberate attention to prompt engineering, even the state-of-the-art LLM-based tools are still deeply unreliable in solving relatively basic physics problems. We speculate that significant efforts will need to be made to engineer (both through advanced prompting techniques but potentially also through model fine-tuning and selective training data) LLM-based chatbots that can serve as reliable tutors for physics. It is hard to predict when this could happen, but given the current pace of development, it is likely that some form of AI-based physics tutoring will materialise in the next couple of years. How successful it will be remains to be seen. As things stand now, we would argue strongly against relying on the current versions of ChatGPT or any other LLM-based chatbots for tutoring purposes.
2. There is some potential to use chatbots, such as ChatGPT, for generating content to be evaluated by humans. This is especially interesting in the context of physics teacher training, as LLMs have been found sometimes to produce statements that resemble student difficulties [17]. On the other hand, LLM can also produce uncanny statements, which are very atypical of real students' reasoning. This presents a challenge for using LLM-based chatbots as models of students, on which future and existing teachers can train to better respond to student difficulties. Additional limitations for such use come from the LLMs' enormous training database, which makes them behave as very confident and knowledgeable on a wide variety of topics – something that can seldom be seen in human students. However, interaction with LLM-based chatbots still holds the potential to be a training tool for the critical assessment of ideas. As Gregorcic and Pendrill [17] have suggested, because LLMs' responses are so well-formulated in terms of grammar and style, this requires the critical reader to "really engage with the substance of their content, instead of focusing on grammatical or stylistic issues" [p.8].
3. Perhaps the most promising way of using LLMs at this point in time is also one that has the potential to mitigate some of the main concerns regarding students' tendency to uncritically accept their output. If an LLM-based chatbot plays the role of a collaborative peer instead of an authority (e.g., a teacher or expert in a given field), this sets up a more productive framing for the critical interpretation of its output. At the same time, such framing also allows students to take advantage of the often-fruitful outputs that these tools can provide. Framing the chatbot as a peer would involve students taking a sceptical stance towards its outputs and employing independent means of evaluating their correctness and usefulness. With recent versions of ChatGPT displaying a "humbler" character, willing to accept criticism and revise its responses (e.g. the example provided in IV.C.6), compared to previous versions [17], this opens the door to dialogic approaches to physics problem solving, a form of collaboration between AI and humans, where LLM-based chatbots can be used as "objects to think with" [81]. As we have demonstrated in this paper, dialogue-based approaches, combined with some basic understanding of LLMs' functioning and prompting techniques, hold the potential to become a useful tool for physics problem solving.

---

[14] The roles proposed here resemble those proposed by Taylor [93] in his seminal work on the use of computers in education.

While tools such as LLM-based chatbots in principle hold the potential to be useful by allowing students to learn more independently, provide them with personalized and real-time responses, and even help reducing teacher work-load [84, 94], they can also be detrimental to learning when used in unproductive ways. Their use involves potential threats to the students' development of deep understanding of subject content and higher-order thinking skills, academic integrity and un-biased world views [84,94].

While the potential benefits and risks are not limited to any specific group, studies point out that students from underserved communities can potentially be more affected, both positively and negatively, by the use of this technology in education [95,96]. One major risk is the potential widening of the digital divide between those with access to LLM-based services and to the know-how of using them, and those who do not have access to them or are not taught how to use them productively. Furthermore, because LLMs are mostly trained on large languages, with English being the most represented, non-speakers of these languages are also in a disadvantaged position [97,98].

Lastly, paying attention and noticing LLMs' biases and stopping the propagation of stereotypes related to who is or can be a physicist is something we should pay special attention to in our efforts to make physics a more inclusive discipline [99].

# VI. CONCLUSION

This paper aims to provide physics educators with some rudimentary insights into the operation of large language models (LLMs). We summarise existing research on the most commonly used LLM-based chatbot's (ChatGPT) performance on physics tasks, pinpoint its strengths and weaknesses, and explain in simplified terms how it functions. We then look at how selected prompting approaches can improve the performance of ChatGPT-4 on introductory physics tasks, focusing on conceptual tasks requiring little to no calculation. We provide several illustrative physics examples, through which we qualitatively explore the effects of different prompting techniques and, by doing this, glean additional insights into the functioning principles of LLMs. Finally, we provide a summary of existing literature on the educational implications of LLMs and supplement it with our own insights and reflections developed through the process of writing this paper. We hope that the paper can serve as a resource for those in the physics education community who wish to develop a better understanding of how these AI-based tools work and how they can be used in physics teaching and learning. We are also hopeful that the paper can inform and facilitate further research at the cross-section of computational linguistics and physics education research.

# REFERENCES

[1] T. P. Tate, S. Doroudi, D. Ritchie, Y. Xu, and M. W. Uci, *Educational Research and AI-Generated Writing: Confronting the Coming Tsunami*, 10.35542/osf.io/4mec3.

[2] Ungdomsbarometern, *Back2School 2023*, https://www.ungdomsbarometern.se/rapportslapp-back2school-2023/?fbclid=IwAR0qrPf9sYyiFzNy0c-f-lZ9APu8sSJPm3b9cOfGZuUfkg95DSCda9-Ob_w.

[3] L. Welding, *Half of College Students Say Using AI on Schoolwork Is Cheating or Plagiarism*, https://www.bestcolleges.com/research/college-students-ai-tools-survey/#:~:text=43%25%20of%20college%20students%20have,college%20students%20in%20our%20survey.

[4] T. B. Brown et al., *Language Models Are Few-Shot Learners*, arXiv:2005.14165.

[5] J. W. Rae et al., *Scaling Language Models: Methods, Analysis & Insights from Training Gopher*, arXiv:2112.11446.

[6] K. Cobbe et al., *Training Verifiers to Solve Math Word Problems*, arXiv:2110.14168.

[7] D. Hendrycks, C. Burns, S. Kadavath, A. Arora, S. Basart, E. Tang, D. Song, and J. Steinhardt, *Measuring Mathematical Problem Solving With the MATH Dataset*, arXiv:2103.03874.

[8] P. Lewis et al., *Retrieval-Augmented Generation for Knowledge-Intensive NLP Tasks*, arXiv:2005.11401.

[9] J. H. Choi, K. E. Hickman, A. Monahan, and D. B. Schwarcz, *ChatGPT Goes to Law School*, SSRN Electron. J. (2023).

[10] W. Geerling, G. D. Mateer, J. Wooten, and N. Damodaran, *ChatGPT Has Mastered the Principles of Economics: Now What?*, SSRN Electron. J. (2023).

[11] H. Nori, N. King, S. M. McKinney, D. Carignan, and E. Horvitz, *Capabilities of GPT-4 on Medical Challenge Problems*, arXiv:2303.13375.

[12] M. Sallam, *The Utility of ChatGPT as an Example of Large Language Models in Healthcare Education, Research and Practice: Systematic Review on the Future Perspectives and Potential Limitations*, 10.1101/2023.02.19.23286155.

[13] S. Frieder, L. Pinchetti, R.-R. Griffiths, T. Salvatori, T. Lukasiewicz, P. C. Petersen, A. Chevalier, and J. Berner, *Mathematical Capabilities of ChatGPT*, arXiv:2301.13867.

[14] P. M. Newton and M. Xiromeriti, *ChatGPT Performance on MCQ Exams in Higher Education. A Pragmatic Scoping Review*, 10.35542/osf.io/sytu3.

[15] OpenAI, *GPT-4 Technical Report*, arXiv:2303.08774.

[16] S. Bubeck et al., *Sparks of Artificial General Intelligence: Early Experiments with GPT-4*, arXiv:2303.12712.

[17] B. Gregorcic and A.-M. Pendrill, *ChatGPT and the Frustrated Socrates*, Phys. Educ. **58**, 035021 (2023).
[18] R. P Dos Santos, *Enhancing Physics Learning with ChatGPT, Bing Chat, and Bard as Agents-to-Think-With: A Comparative Case Study*, SSRN Electron. J. (2023).
[19] G. Kortemeyer, *Could an Artificial-Intelligence Agent Pass an Introductory Physics Course?*, Phys. Rev. Phys. Educ. Res. **19**, 010132 (2023).
[20] D. Hestenes, M. Wells, and G. Swackhamer, *Force Concept Inventory*, Phys. Teach. **30**, 141 (1992).
[21] C. G. West, *AI and the FCI: Can ChatGPT Project an Understanding of Introductory Physics?*, arXiv:2303.01067.
[22] C. G. West, *Advances in Apparent Conceptual Physics Reasoning in GPT-4*, arXiv:2303.17012.
[23] W. Yeadon, O.-O. Inyang, A. Mizouri, A. Peach, and C. P. Testrow, *The Death of the Short-Form Physics Essay in the Coming AI Revolution*, Phys. Educ. **58**, 035027 (2023).
[24] W. Yeadon and T. Hardy, *The Impact of AI in Physics Education: A Comprehensive Review from GCSE to University Levels*, arXiv:2309.05163.
[25] W. Yeadon and D. P. Halliday, *Exploring Durham University Physics Exams with Large Language Models*, arXiv:2306.15609.
[26] E. Etkina, G. Planinsic, and A. Van Heuvelen, *College Physics: Explore and Apply*, 2nd Editio (Pearson, New York, NY, 2019).
[27] L. Ouyang et al., *Training Language Models to Follow Instructions with Human Feedback*, arXiv:2203.02155.
[28] Y. Wang, W. Zhong, L. Li, F. Mi, X. Zeng, W. Huang, L. Shang, X. Jiang, and Q. Liu, *Aligning Large Language Models with Human: A Survey*, arXiv:2307.12966.
[29] A. Vaswani, N. Shazeer, N. Parmar, J. Uszkoreit, L. Jones, A. N. Gomez, L. Kaiser, and I. Polosukhin, *Attention Is All You Need*, arXiv:1706.03762.
[30] M. Shanahan, *Talking About Large Language Models*, arXiv:2212.03551.
[31] Z. Ji, N. Lee, R. Frieske, T. Yu, D. Su, Y. Xu, E. Ishii, Y. J. Bang, A. Madotto, and P. Fung, *Survey of Hallucination in Natural Language Generation*, ACM Comput. Surv. **55**, 1 (2023).
[32] F. Petroni, P. Lewis, A. Piktus, T. Rocktäschel, Y. Wu, A. H. Miller, and S. Riedel, *How Context Affects Language Models' Factual Predictions*, arXiv:2005.04611.
[33] L. Floridi and M. Chiriatti, *GPT-3: Its Nature, Scope, Limits, and Consequences*, Minds Mach. **30**, 681 (2020).
[34] A. Borji, *A Categorical Archive of ChatGPT Failures*, arXiv:2302.03494.
[35] C. K. Lo, *What Is the Impact of ChatGPT on Education? A Rapid Review of the Literature*, Educ. Sci. **13**, 410 (2023).
[36] V. Thakur, *Unveiling Gender Bias in Terms of Profession Across LLMs: Analyzing and Addressing Sociological Implications*, arXiv:2307.09162.
[37] H. Kotek, R. Dockum, and D. Q. Sun, *Gender Bias and Stereotypes in Large Language Models*, in *Proceedings of The ACM Collective Intelligence Conference* (2023), pp. 12–24.
[38] J. A. Omiye, J. C. Lester, S. Spichak, V. Rotemberg, and R. Daneshjou, *Large Language Models Propagate Race-Based Medicine*, Npj Digit. Med. **6**, 195 (2023).
[39] X. Fang, S. Che, M. Mao, H. Zhang, M. Zhao, and X. Zhao, *Bias of AI-Generated Content: An Examination of News Produced by Large Language Models*, arXiv:2309.09825.
[40] K. Khandelwal, M. Tonneau, A. M. Bean, H. R. Kirk, and S. A. Hale, *Casteist but Not Racist? Quantifying Disparities in Large Language Model Bias between India and the West*, arXiv:2309.08573.
[41] F. Motoki, V. Pinho Neto, and V. Rodrigues, *More Human than Human: Measuring ChatGPT Political Bias*, Public Choice (2023).
[42] I. O. Gallegos, R. A. Rossi, J. Barrow, M. M. Tanjim, S. Kim, F. Dernoncourt, T. Yu, R. Zhang, and N. K. Ahmed, *Bias and Fairness in Large Language Models: A Survey*, arXiv:2309.00770.
[43] Y. Bang et al., *A Multitask, Multilingual, Multimodal Evaluation of ChatGPT on Reasoning, Hallucination, and Interactivity*, arXiv:2302.04023.
[44] T. Chiang, *ChatGPT Is a Blurry JPEG of the Web*, https://www.newyorker.com/tech/annals-of-technology/chatgpt-is-a-blurry-jpeg-of-the-web.

[45] X. Shen, Z. Chen, M. Backes, and Y. Zhang, *In ChatGPT We Trust? Measuring and Characterizing the Reliability of ChatGPT*, arXiv:2304.08979.

[46] E. Davis, *Mathematics, Word Problems, Common Sense, and Artificial Intelligence*, arXiv:2301.09723.

[47] S. Wolfram, *ChatGPT Gets Its "Wolfram Superpowers"!*, https://writings.stephenwolfram.com/2023/03/chatgpt-gets-its-wolfram-superpowers/?fbclid=IwAR1oyIMHMhH4Oi9yua7HkjOUGS3mznGDgewd9nZRnf1PyuHkflgMx2Cl4dg.

[48] D. Hestenes, *Modeling Games in the Newtonian World*, Am. J. Phys. **60**, 732 (1992).

[49] E. F. Redish, *Implications of Cognitive Studies for Teaching Physics*, Am. J. Phys. **62**, 796 (1994).

[50] R. Goodwins, *ChatGPT Has Mastered the Confidence Trick, and That's a Terrible Look for AI*, https://www.theregister.com/2022/12/12/chatgpt_has_mastered_the_confidence/.

[51] V. Talanquer, *Interview with the Chatbot: How Does It Reason?*, J. Chem. Educ. acs.jchemed.3c00472 (2023).

[52] J. Rudolph, S. Tan, and S. Tan, *ChatGPT: Bullshit Spewer or the End of Traditional Assessments in Higher Education?*, J. Appl. Learn. Teach. **6**, (2023).

[53] J. D. Zamfirescu-Pereira, R. Y. Wong, B. Hartmann, and Q. Yang, *Why Johnny Can't Prompt: How Non-AI Experts Try (and Fail) to Design LLM Prompts*, in *Proceedings of the 2023 CHI Conference on Human Factors in Computing Systems* (ACM, Hamburg Germany, 2023), pp. 1–21.

[54] M. Salah, H. Alhalbusi, M. M. Ismail, and F. Abdelfattah, *Chatting with ChatGPT: Decoding the Mind of Chatbot Users and Unveiling the Intricate Connections between User Perception, Trust and Stereotype Perception on Self-Esteem and Psychological Well-Being*, Curr. Psychol. (2023).

[55] J. White, S. Hays, Q. Fu, J. Spencer-Smith, and D. C. Schmidt, *ChatGPT Prompt Patterns for Improving Code Quality, Refactoring, Requirements Elicitation, and Software Design*, arXiv:2303.07839.

[56] J. White, Q. Fu, S. Hays, M. Sandborn, C. Olea, H. Gilbert, A. Elnashar, J. Spencer-Smith, and D. C. Schmidt, *A Prompt Pattern Catalog to Enhance Prompt Engineering with ChatGPT*, arXiv:2302.11382.

[57] T. Kojima, S. S. Gu, M. Reid, Y. Matsuo, and Y. Iwasawa, *Large Language Models Are Zero-Shot Reasoners*, arXiv:2205.11916.

[58] S. R. Bowman, *Eight Things to Know about Large Language Models*, arXiv:2304.00612.

[59] K. Valmeekam, A. Olmo, S. Sreedharan, and S. Kambhampati, *Large Language Models Still Can't Plan (A Benchmark for LLMs on Planning and Reasoning about Change)*, arXiv:2206.10498.

[60] B. Edwards, *NASCAR Driver Stuns Racing World with a Move Learned from Nintendo GameCube*, https://arstechnica.com/gaming/2022/11/nascar-driver-stuns-racing-world-with-a-move-learned-from-nintendo-gamecube/.

[61] B. Xu, A. Yang, J. Lin, Q. Wang, C. Zhou, Y. Zhang, and Z. Mao, *ExpertPrompting: Instructing Large Language Models to Be Distinguished Experts*, arXiv:2305.14688.

[62] F. Shi, X. Chen, K. Misra, N. Scales, D. Dohan, E. Chi, N. Schärli, and D. Zhou, *Large Language Models Can Be Easily Distracted by Irrelevant Context*, arXiv:2302.00093.

[63] J. Wei et al., *Emergent Abilities of Large Language Models*, arXiv:2206.07682.

[64] J. Wei, X. Wang, D. Schuurmans, M. Bosma, B. Ichter, F. Xia, E. Chi, Q. Le, and D. Zhou, *Chain-of-Thought Prompting Elicits Reasoning in Large Language Models*, arXiv:2201.11903.

[65] D. Kahneman and S. Frederick, *Representativeness Revisited: Attribute Substitution in Intuitive Judgment*, in *Heuristics and Biases*, edited by T. Gilovich, D. Griffin, and D. Kahneman, 1st ed. (Cambridge University Press, 2002), pp. 49–81.

[66] S. C. Bellini-Leite, *Analytic Thinking (Type 2 or "System 2") for Large Language Models: Using Psychology to Address Hallucination and Reliability Issues*, 10.35542/osf.io/4mec3.

[67] OpenAI, *Models*, https://platform.openai.com/docs/models/overview.

[68] Q. Dong, L. Li, D. Dai, C. Zheng, Z. Wu, B. Chang, X. Sun, J. Xu, L. Li, and Z. Sui, *A Survey on In-Context Learning*, arXiv:2301.00234.
31

[69] X. Wang, J. Wei, D. Schuurmans, Q. Le, E. Chi, S. Narang, A. Chowdhery, and D. Zhou, *Self-Consistency Improves Chain of Thought Reasoning in Language Models*, arXiv:2203.11171.
[70] J. Long, *Large Language Model Guided Tree-of-Thought*, arXiv:2305.08291.
[71] S. Yao, D. Yu, J. Zhao, I. Shafran, T. L. Griffiths, Y. Cao, and K. Narasimhan, *Tree of Thoughts: Deliberate Problem Solving with Large Language Models*, arXiv:2305.10601.
[72] E. Y. Chang, *Prompting Large Language Models With the Socratic Method*, in *2023 IEEE 13th Annual Computing and Communication Workshop and Conference (CCWC)* (IEEE, Las Vegas, NV, USA, 2023), pp. 0351–0360.
[73] T. Liang, Z. He, W. Jiao, X. Wang, Y. Wang, R. Wang, Y. Yang, Z. Tu, and S. Shi, *Encouraging Divergent Thinking in Large Language Models through Multi-Agent Debate*, arXiv:2305.19118.
[74] W. X. Zhao et al., *A Survey of Large Language Models*, arXiv:2303.18223.
[75] B. Min, H. Ross, E. Sulem, A. P. B. Veyseh, T. H. Nguyen, O. Sainz, E. Agirre, I. Heintz, and D. Roth, *Recent Advances in Natural Language Processing via Large Pre-Trained Language Models: A Survey*, ACM Comput. Surv. **56**, 1 (2023).
[76] W. Holmes, M. Bialik, and C. Fadel, *Promises and Implications for Teaching and Learning.*, in *Artificial Intelligence in Education.* (Center for Curriculum Redesign, Boston, 2019).
[77] O. E. Chinonso, A. M.-E. Theresa, and T. C. Aduke, *ChatGPT for Teaching, Learning and Research: Prospects and Challenges*, Glob. Acad. J. Humanit. Soc. Sci. **5**, 33 (2023).
[78] E. Kasneci et al., *ChatGPT for Good? On Opportunities and Challenges of Large Language Models for Education*, Learn. Individ. Differ. **103**, 102274 (2023).
[79] Md. M. Rahman and Y. Watanobe, *ChatGPT for Education and Research: Opportunities, Threats, and Strategies*, Appl. Sci. **13**, 5783 (2023).
[80] P. P. Ray, *ChatGPT: A Comprehensive Review on Background, Applications, Key Challenges, Bias, Ethics, Limitations and Future Scope*, Internet Things Cyber-Phys. Syst. **3**, 121 (2023).
[81] M. A. R. Vasconcelos and R. P. Dos Santos, *Enhancing STEM Learning with ChatGPT and Bing Chat as Objects to Think with: A Case Study*, Eurasia J. Math. Sci. Technol. Educ. **19**, em2296 (2023).
[82] M. A. AlAfnan, Samira Dishari, Marina Jovic, and Koba Lomidze, *ChatGPT as an Educational Tool: Opportunities, Challenges, and Recommendations for Communication, Business Writing, and Composition Courses*, J. Artif. Intell. Technol. (2023).
[83] J. Crawford, M. Cowling, Central Queensland University, Australia, K.-A. Allen, and Monash University, Australia, *Leadership Is Needed for Ethical ChatGPT: Character, Assessment, and Learning Using Artificial Intelligence (AI)*, J. Univ. Teach. Learn. Pract. **20**, (2023).
[84] T. Adiguzel, M. H. Kaya, and F. K. Cansu, *Revolutionizing Education with AI: Exploring the Transformative Potential of ChatGPT*, Contemp. Educ. Technol. **15**, ep429 (2023).
[85] D. Baidoo-Anu and L. Owusu Ansah, *Education in the Era of Generative Artificial Intelligence (AI): Understanding the Potential Benefits of ChatGPT in Promoting Teaching and Learning*, SSRN Electron. J. (2023).
[86] G. Cooper, *Examining Science Education in ChatGPT: An Exploratory Study of Generative Artificial Intelligence*, J. Sci. Educ. Technol. **32**, 444 (2023).
[87] M. Halaweh, *ChatGPT in Education: Strategies for Responsible Implementation*, Contemp. Educ. Technol. **15**, ep421 (2023).
[88] C. W. Okonkwo and A. Ade-Ibijola, *Chatbots Applications in Education: A Systematic Review*, Comput. Educ. Artif. Intell. **2**, 100033 (2021).
[89] A. Shoufan, *Exploring Students' Perceptions of ChatGPT: Thematic Analysis and Follow-Up Survey*, IEEE Access **11**, 38805 (2023).
[90] L. Krupp, S. Steinert, M. Kiefer-Emmanouilidis, K. E. Avila, P. Lukowicz, J. Kuhn, S. Küchemann, and J. Karolus, *Unreflected Acceptance -- Investigating the Negative Consequences of ChatGPT-Assisted Problem Solving in Physics Education*, arXiv:2309.03087.
[91] M. N. Dahlkemper, S. Z. Lahme, and P. Klein, *How Do Physics Students Evaluate Artificial Intelligence Responses on Comprehension Questions? A Study on the Perceived Scientific Accuracy and Linguistic Quality of ChatGPT*, Phys. Rev. Phys. Educ. Res. **19**, 010142 (2023).
[92] P. Bitzenbauer, *ChatGPT in Physics Education: A Pilot Study on Easy-to-Implement Activities*, Contemp. Educ. Technol. **15**, ep430 (2023).

[93] R. Taylor, *The Computer in the School: Tutor, Tool, Tutee* (Teachers College Press, Totowa, NJ, 1980).
[94] M. Farrokhnia, S. K. Banihashem, O. Noroozi, and A. Wals, *A SWOT Analysis of ChatGPT: Implications for Educational Practice and Research*, Innov. Educ. Teach. Int. 1 (2023).
[95] P. P. Ray, *ChatGPT: A Comprehensive Review on Background, Applications, Key Challenges, Bias, Ethics, Limitations and Future Scope*, Internet Things Cyber-Phys. Syst. **3**, 121 (2023).
[96] D. Mhlanga, *ChatGPT in Education: Exploring Opportunities for Emerging Economies to Improve Education with ChatGPT*, SSRN Electron. J. (2023).
[97] P. Dave, *ChatGPT Is Cutting Non-English Languages Out of the AI Revolution*, https://www.wired.com/story/chatgpt-non-english-languages-ai-revolution/.
[98] S. Nguyen and C. Anderson, *Do All Minority Languages Look the Same to GPT-3? Linguistic (Mis)Information in a Large Language Model*, Proc. Soc. Comput. Linguist. **6**, 400 (2023).
[99] APS, *Statement on Diversity in Physics*, https://www.aps.org/policy/statements/08_2.cfm.